\documentclass[journal=jctcce,manuscript=article,layout=twocolumn]{achemso}

\usepackage[version=3]{mhchem} 
\usepackage{booktabs}
\usepackage{subcaption}
\usepackage{threeparttable}

\DeclareMathOperator\erf{erf}

\author{Benjamin B. Ye}
\author{Shensheng Chen}
\author{Zhen-Gang Wang}
\email{zgw@caltech.edu}
\phone{+1 (626) 395-4647}
\fax{+1 (626) 568-8743}
\affiliation[California Institute of Technology]
{Division of Chemistry and Chemical Engineering, California Institute 
 of Technology, Pasadena, California, USA}

\title{GCMe: Efficient implementation of the Gaussian core model with smeared electrostatic interactions for molecular dynamics simulations of soft matter systems}

\abbreviations{}
\keywords{}

\begin{document}


\begin{abstract}
In recent years, molecular dynamics (MD) simulations have emerged as an essential tool for understanding the structure, dynamics, and phase behavior in charged soft matter systems. To explore phenomena across greater length and time scales in MD simulations, molecules are often coarse-grained for better computational performance. However, commonly-used force fields represent particles as hard-core interaction centers with point charges, which often overemphasizes the packing effect and short-range electrostatics, especially in systems with bulky deformable organic molecules and systems with strong coarse-graining. This underscores the need for an efficient soft-core model to physically capture the effective interactions between coarse-grained particles. To this end, we implement a soft-core model uniting the Gaussian core model with smeared electrostatic interactions that is phenomenologically equivalent to recent theoretical models. We first parameterize it generically using water as the model solvent. Then, we benchmark its performance in the OpenMM toolkit for different boundary conditions to highlight a computational speedup of up to $34\times$ compared to commonly used force fields and existing implementations. Finally, we demonstrate its utility by investigating how boundary polarizability affects the adsorption behavior of a polyelectrolyte solution on perfectly conducting and nonmetal electrodes.
\end{abstract}

\section*{Introduction}
In recent years, computer simulations have become an essential tool for gaining physical insights into intriguing phenomena in polymeric and soft matter systems. Notably, molecular dynamics (MD) simulations with atomistic force fields, which can accurately reproduce the structural, thermodynamic, and dynamic properties of materials,\cite{gartner_modeling_2019,joshi_review_2021} have been used to understand molecular details and observe dynamic processes in those systems. However, even with the rapid advancement of computing power over the past few decades, atomistic MD simulations are often unable to capture phenomena that require considerable system sizes or extensive equilibrium periods due to the high computational costs of evaluating the interactions between the numerous atoms in large systems.

This limitation has motivated the development of several coarse-grained simulation methods that improve computational performance by simplifying the representation of atomistic systems. A widely used approach is coarse-grained MD simulation, in which groups of atoms or molecules are mapped onto simple, featureless beads to decrease the number of degrees of freedom.\cite{nielsen_coarse_2004,ingolfsson_power_2014} By carefully determining force field parameters using bottom-up (structure-based) and/or top-down (thermodynamics-based) approaches, coarse-grained MD models can faithfully replicate key experimental data despite the loss of detail from the coarse-graining.\cite{brini_systematic_2013} Another coarse-graining strategy is to convert the particle-based representation into a field-based representation through identify transformations. Field-theoretic simulations (FTSs) are particularly well-suited for efficient exploration of mesoscopic length-scale phenomena in high-density systems and polymeric liquids with high molecular weights.\cite{fredrickson_field-theoretic_2002,lee_complex_2008,matsen_field_2020,fredrickson_field-theoretic_2023,villet_efficient_2014,delaney_recent_2016,lequieu_combining_2023} However, FTS does not provide dynamic information and the field variables do not directly inform the molecular-level structure.

Within the common coarse-grained MD models,\cite{jorgensen_optimized_1984,jorgensen_optimized_1985,jorgensen_optimized_1986,souza_martini_2021} such as the Kremer--Grest model for polymeric systems,\cite{kremer_dynamics_1990} the nonbonded interactions are generally treated using a combination of 12-6 Lennard-Jones (LJ) and point-charge Coulomb potentials. While these potentials are ubiquitous in MD simulations, they can be unsuitable for modeling organic molecules that can deform and interpenetrate, or systems that have undergone strong coarse-graining. The LJ potential is inherently a hard-core model because it has a divergence in potential energy when the particle centers come into contact. This places a tight upper bound on the simulation timestep, and can lead to a caging effect that severely hinders the diffusion of particles or even unphysical bulk liquid--solid phase transitions at high geometric packing.\cite{frisch_hard-sphere_1965} Similarly, the Coulomb potential cannot capture the charge delocalization in bulky organic ions because it assigns the entirety of a particle's charge to its interaction center. As such, softer interactions are better suited to reflect the less pronounced packing effects and short-range electrostatics in coarse-grained soft matter systems.

Dissipative particle dynamics (DPD),\cite{hoogerbrugge_simulating_1992,espanol_statistical_1995} a mesoscale simulation method commonly used to model fluidic systems, addresses the shortcomings of hard-core models by treating excluded volume interactions using a soft-core potential.\cite{groot_dissipative_1997} To model charged particles, the charges are smeared onto an arbitrarily-sized grid to prevent unbreakable ion pairs.\cite{groot_electrostatic_2003} While the DPD potential is well-suited for modeling the soft interactions between coarse-grained bulky molecules or even entire liquid elements, its quadratic form is not easily amenable to theoretical analysis. Perhaps more importantly, there is no analytically closed form for the electrostatic potential, so it is usually evaluated on CPU without Ewald summation in the MD software that supports DPD. This makes the evaluation of the electrostatic interactions much slower than the GPU-accelerated implementations commonly used in LJ-based charged systems.

Therefore, there is a need for a more physical and efficient GPU-enabled coarse-grained MD model that can simulate large-scale soft matter systems over long time frames. To this end, we implement a coarse-grained MD model that is motivated by and compatible with previous Gaussian soft-core models, such as the Gaussian overlap model by Berne and Pechukas\cite{berne_gaussian_1972} and the Gaussian core model (GCM) by Stillinger.\cite{stillinger_phase_1976,prestipino_phase_2005,ruppeiner_thermodynamic_2021} These types of models have been used recently to study the liquid--liquid phase separation and dynamics in polymer solutions and melts.\cite{guenza_cooperative_2001,guenza_intermolecular_2002,villet_efficient_2014,delaney_recent_2016,mccarty_complete_2019,shen_learning_2020} In particular, Jedlinska et al.\cite{jedlinska_matildaft_2023} and Lequieu\cite{lequieu_combining_2023} have employed Gaussian-distributed particle densities in their multi-representation simulation approaches to convert between particle- and field-based representations. Our model, henceforth referred to as GCMe, combines the GCM with smeared electrostatic interactions from Gaussian charge distributions.\cite{warren_screening_2013,kiss_efficient_2014,eslami_gaussian_2019} This combination maintains the soft nature of DPD and its advantages, can model the charge smearing in large ions, and still has simple analytically tractable expressions for the interaction potentials. In this work, we first establish the thermodynamic basis of GCMe and highlight the computational speedup possible with our GCMe implementation in the high-performance OpenMM toolkit for different types of system boundaries. Then, as an example of a practical application of GCMe, we examine the effects of the electrode polarizability on the adsorption of the charged species in polyelectrolyte solutions.

\section*{Model and methods}
\subsection*{Gaussian core model}
In GCMe, featureless spherical particles with smeared charges interact via centrosymmetric pair potentials. A charged particle $i$ has Gaussian-distributed mass density
\begin{equation} \label{eq:mass_distribution}
  \rho_{i,\,m}(\mathbf{r})=\left(\frac{3}{2\pi\sigma_i^2}\right)^{3/2}\exp{\left(-\frac{3}{2\sigma_i^2}(\mathbf{r}-\mathbf{r}_i)^2\right)}
\end{equation}
and charge density
\begin{equation} \label{eq:charge_distribution}
  \rho_{i,\,q}(\mathbf{r})=\frac{z_ie}{(2a_i^2)^{3/2}}\exp\left(-\frac{\pi}{2a_i^2}(\mathbf{r}-\mathbf{r}_i)^2\right),
\end{equation}
where $\sigma_i$ and $a_i$ are the mass and electrostatic smearing radii, respectively, $z_i$ is the charge number, and $e$ is the elementary charge. These Gaussian distributions have been used in theoretical models\cite{guenza_cooperative_2001,guenza_intermolecular_2002,wang_fluctuation_2010,villet_efficient_2014,delaney_recent_2016,mccarty_complete_2019,shen_learning_2020} recently because they have conveniently defined Fourier transforms and approach the Dirac delta function, the limit of a point charge, as the smearing radii approach zero, and the particular form of the charge density distribution in Eq. \ref{eq:charge_distribution} reproduces the Born energy for the self-energy of an ion $i$.\cite{wang_fluctuation_2010}

The pair interaction potentials for Eqs. \ref{eq:mass_distribution} and \ref{eq:charge_distribution} are analytically solvable integrals, and the derivations can be found in the Supporting Information (SI). The excluded volume interaction between two particles $i$ and $j$ is given by
\begin{equation} \label{eq:u_ex}
  u_\mathrm{ex}(r_{ij})=A_{ij}\left(\frac{3}{2\pi\sigma_{ij}^2}\right)^{3/2}\exp{\left(-\frac{3}{2\sigma_{ij}^2}r_{ij}^2\right)},
\end{equation}
where $\sigma_{ij}=\sqrt{\sigma_i^2+\sigma_j^2}$, $r_{ij}$ is the separation distance, and $A_{ij}$ specifies the strength of the excluded volume interactions. Since Eq. \ref{eq:u_ex} has the form of a Gaussian interaction potential, the local excluded volume interaction between two smeared mass densities is equivalent to a Gaussian interaction between two point particles.

The smeared electrostatic interaction has the form
\begin{equation} \label{eq:u_elec}
  u_\mathrm{elec}(r_{ij})=\frac{z_iz_je^2}{4\pi\varepsilon_0\varepsilon_\mathrm{r}r_{ij}}\erf\left(\frac{\pi^{1/2}}{2^{1/2}a_{ij}}r_{ij}\right),
\end{equation}
where $a_{ij}=\sqrt{a_i^2+a_j^2}$, $\varepsilon_0$ is the vacuum permittivity, and $\varepsilon_\mathrm{r}$ is the relative permittivity. Unlike the point-charge Coulomb potential, the smeared electrostatic potential remains finite even as the separation distance tends to zero. Importantly, the self-energy of an ion $i$ ($r_{ii}\rightarrow0$) can now be defined and is simply the Born energy $z_i^2e^2/(8\pi\varepsilon_0\varepsilon_\mathrm{r}a)$. For two particles that come into contact ($r_{ij}\rightarrow0$), the smeared electrostatic potential reaches an asymptotic value of $z_iz_je^2/(2^{3/2}\pi\varepsilon_0\varepsilon_\mathrm{r}a_{ij})$. To prevent ions from collapsing on top of each other, the excluded volume repulsion must overcome the electrostatic attraction between two oppositely charged particles, so the relationship
\begin{equation}
  \frac{A_{ij}}{\sigma_{ij}^3k_\mathrm{B}T}\gg\frac{4\pi^{3/2}z_iz_j\lambda_\mathrm{B}}{3^{3/2}a_{ij}},
\end{equation}
where $\lambda_\mathrm{B}=e^2/(4\pi\varepsilon_0\varepsilon_\mathrm{r}k_\mathrm{B}T)$ is the Bjerrum length, must hold.

Another advantage of the smeared electrostatic potential is that it can still be efficiently evaluated in reciprocal space using three-dimensional Ewald summation.\cite{ewald_berechnung_1921} The particle-mesh Ewald method for the smeared electrostatic potential splits the total coulombic energy into four parts,
\begin{align} \label{eq:ewald}
  \begin{split}
    U_\mathrm{elec}&=\frac{e^2}{8\pi\varepsilon_0\varepsilon_\mathrm{r}}\sum_\mathbf{n}\sum_{i,j}^N\frac{z_iz_j}{r_{ij,\,\mathbf{n}}}\erf{\left(\frac{\pi^{1/2}}{2^{1/2}a_{ij}}r_{ij,\,\mathbf{n}}\right)}\\
    &=U_\mathrm{elec,\,real}+U_\mathrm{elec,\,recip}+U_\mathrm{elec,\,corr}+U_\mathrm{elec,\,self}\\
    &=\frac{e^2}{8\pi\varepsilon_0\varepsilon_\mathrm{r}}\sum_\mathbf{n}^\prime\sum_{i,j}^N\frac{z_iz_j}{r_{ij,\,\mathbf{n}}}\biggl[\erf{\left(\frac{\pi^{1/2}}{2^{1/2}a_{ij}}r_{ij,\,\mathbf{n}}\right)}\\
    &\qquad\qquad\qquad\qquad\qquad-\erf\left(\frac{r_{ij,\,\mathbf{n}}}{2^{1/2}G}\right)\biggr]\\
    &\quad+\frac{1}{2\varepsilon_0\varepsilon_\mathrm{r}V}\sum_{\mathbf{k}\neq\mathbf{0}}\frac{\exp{\left(-G^2k^2/2\right)}}{k^2}|S(\mathbf{k})|^2\\
    &\quad-\frac{e^2}{4\pi(2\pi)^{1/2}\varepsilon_0\varepsilon_\mathrm{r}G}\sum_i^Nz_i^2\\
    &\quad+\frac{e^2}{8\pi\varepsilon_0\varepsilon_\mathrm{r}}\sum_i^N\frac{z_i^2}{a_i},
  \end{split}
\end{align}
where $\mathbf{n}$ is the cell-coordinate vector specifying the position of a periodic cell, the prime ($\prime$) in the sum in the $U_\mathrm{elec,\,real}$ term indicates that terms where $i=j$ are omitted for $\mathbf{n}=\mathbf{0}$, $G$ is the width of the Gaussian charge distributions introduced to screen the smeared charges, $\mathbf{k}$ and $k=|\mathbf{k}|$ are the reciprocal lattice vector and its magnitude, respectively, and $S(\mathbf{k})=e\sum_j^N z_j\exp{(i\mathbf{k}\cdot\mathbf{r}_j)}$.\cite{gingrich_ewald_2010,coslovich_ultrasoft_2011,kiss_efficient_2014,eslami_gaussian_2019} The screening charge distributions are not strictly necessary, as the evaluation of the position-dependent terms in Eq. \ref{eq:ewald} can be carried out entirely in reciprocal space by choosing $G=a_{ij}/\pi^{1/2}$, but are generally included to improve computational efficiency by moving part of the calculation effort to the real space.\cite{coslovich_ultrasoft_2011,eslami_gaussian_2019}

In Eq. \ref{eq:ewald}, the $U_\mathrm{elec,\,recip}$ and $U_\mathrm{elec,\,corr}$ terms are equivalent to those found in the standard Ewald summation for the point-charge Coulomb potential.\cite{darden_particle_1993,essmann_smooth_1995,toukmaji_ewald_1996} The $U_\mathrm{Coul,\,recip}$ term captures the coulombic energy of point charges (carrying the same charges as the smeared charges) interacting with equal but opposite charge distributions compensating the screening charge distributions in reciprocal space after the cutoff, and the $U_\mathrm{Coul,\,corr}$ term is a constant term that corrects for the self-interactions between the point charges and the compensating charge distributions spuriously included in $U_\mathrm{Coul,\,recip}$. 

The only differences between the Ewald summations for smeared charges and point charges are manifested in the $U_\mathrm{elec,\,real}$ term and an additional constant $U_\mathrm{elec,\,self}$ term that accounts for the self-energies of the ions.\cite{gingrich_ewald_2010} The $U_\mathrm{elec,\,real}$ term evaluates short-range electrostatic interactions within a designated cutoff in real space. Inside the brackets, the first term corresponds to the interaction between two smeared charges $i$ and $j$ (instead of two point charges), while the second term accounts for the interaction between a point charge and the screening charge distribution for charge $j$ that cancels out the compensating interaction in $U_\mathrm{Coul,\,recip}$.

As such, existing mesh-based schemes for evaluating the standard Ewald summation can be used for Eq. \ref{eq:ewald} by simply modifying the real-space contribution and including the self-energies. Schematics depicting and comparing the real-space, reciprocal-space, correction, and self-energy contributions to the Ewald summations for point charges and GCMe smeared charges can be found in Fig. SI in the SI.

\subsection*{Parametrization}
An important criteria for coarse-grained models is that the thermodynamics of the fluid, which is described by the fluctuations in the system, should be reproduced faithfully. After selecting appropriate $\sigma_i$ and $a_i$ values for a system of interest, there is only one undefined GCMe parameter, the repulsion parameter $A_{ij}$. In principle, $A_{ij}$ can be determined precisely via systematic structure- or thermodynamics-based coarse-graining procedures.\cite{brini_systematic_2013} However, in this work, we aim to simply model generic soft matter systems in a manner consistent with how equivalent DPD simulations would be configured. Following the formulation of the DPD method,\cite{groot_dissipative_1997,groot_mesoscopic_2001,groot_electrostatic_2003} we also parameterize GCMe by associating the value of $A_{ij}$ with the compressibility of water, a commonly-used solvent in soft matter systems.

To connect $A_{ij}$ to the compressibility, we start with an expression relating the pressure $p$ to the simulation number density $\rho$. Using the virial theorem and the additivity of the pairwise forces, the pressure can be determined using
\begin{equation} \label{eq:pres_virial}
  \begin{split}
    p&=\rho k_\mathrm{B}T+\frac{1}{3V}\left\langle\sum_{i<j}(\mathbf{r}_i-\mathbf{r}_j)\cdot\mathbf{f}_i\right\rangle\\
    &=\rho k_\mathrm{B}T+\frac{2\pi}{3}\rho^2\int_0^\infty r^3f_{ij}(r)g_{ij}(r)\,dr,
  \end{split}
\end{equation}
where $k_\mathrm{B}T$ is the thermal energy scale, $V$ is the system volume, $\mathbf{f}_i$ and $\mathbf{r}_i$ are the force acting on and the position of particle $i$, and $f_{ij}$ and $g_{ij}$ are the pairwise GCMe force and radial distribution function between particles $i$ and $j$. For soft sphere models, the main contributions to the pressure come from the leading-order $\rho$ terms,\cite{groot_dissipative_1997} so an approximation for the equation of state that holds at sufficiently high number densities is
\begin{equation} \label{eq:pres_approx}
  p\approx\rho k_\mathrm{B}T+\omega A_{ij}\rho^2,
\end{equation}
where $\omega$ is a dimensionless scaling constant. With Eq. \ref{eq:pres_approx}, the dimensionless compressibility $\kappa^{-1}$ is related to $A_{ij}$ via
\begin{equation} \label{eq:inverse_kappa}
  \begin{split}
    \kappa^{-1}&=\frac{1}{nk_\mathrm{B}T\kappa_T}=\frac{1}{k_\mathrm{B}T}\left(\frac{\partial p}{\partial\rho}\right)_T\left(\frac{\partial\rho}{\partial n}\right)_T\\
    &=\frac{1}{N_\mathrm{m}}\left(1+\frac{2\omega A_{ij}\rho}{k_\mathrm{B}T}\right),
  \end{split}
\end{equation}
where $n$ is the number density of water molecules, $\kappa_T$ is the isothermal compressibility, and $N_\mathrm{m} \equiv (\partial n / \partial \rho)_T$ is a real-space renormalization factor, or the number of water molecules represented by each simulation particle. 

To determine $\omega$ and obtain the equation of state, we conducted a series of MD simulations in the isothermal--isobaric ($NpT$) ensemble at room temperature $T=300$ $\mathrm{K}$ and varying $p$ and $A_{ij}$ values. The cubic systems were initialized with $N = 10,000$ randomly placed particles with size $d=2\sigma=0.275$ $\mathrm{nm}$ and mass $m=18.02$ $\mathrm{g/mol}$. For each simulation, energy minimization and system equilibration were performed over $1\times10^7$ timesteps of step size $t=0.01\tau$, where $\tau=\sqrt{md^2/(N_\mathrm{A}k_\mathrm{B}T)}\approx0.739$ $\mathrm{ps}$ is the intrinsic time scale and $N_\mathrm{A}$ is the Avogadro constant, and data were collected over an additional $4 \times 10^7$ timesteps, which is on the order of $300$ $\mathrm{ns}$.

\begin{figure}[ht]
  \centering
  \includegraphics[width=3.25in]{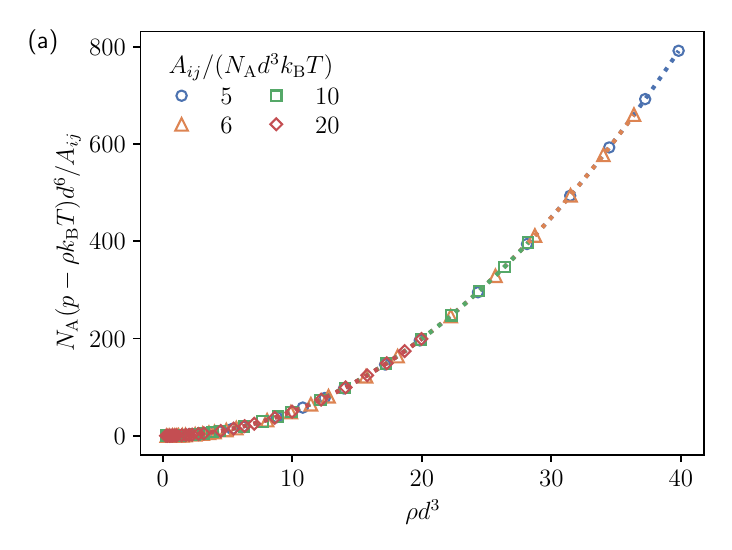}
  \includegraphics[width=3.25in]{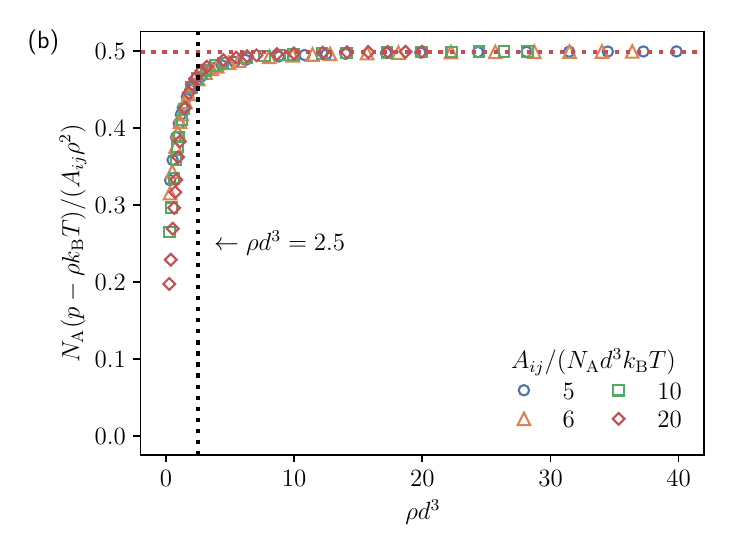}
  \caption{(a) $A_{ij}$- and (b) $A_{ij} \rho^2$-normalized excess pressure $p - \rho k_\mathrm{B} T$ as functions of the number density $\rho$ and repulsion parameter $A_{ij}$. The dotted lines in (a) show the best quadratic fits. The horizontal dotted lines in (b) specify the asymptotic $\omega$ values, while the vertical dotted line indicate the optimal number density $\rho$.}
  \label{fig:water_p_excess_vs_rho}
\end{figure}

By plotting the excess pressure (or the non-ideal contributions) normalized by $A_{ij}$, Fig. \ref{fig:water_p_excess_vs_rho}a shows that all systems fall on a master curve, indicating a simple scaling relation. When the excess pressure is normalized by $\rho^2$, it levels off to an asymptotic value of $\omega=0.499$ at large $\rho$ in Fig. \ref{fig:water_p_excess_vs_rho}b, confirming that the main contribution to the excess pressure is from the $\rho^2$ term and that Eq. \ref{eq:pres_approx} is valid. With $\omega=0.499$ and a dimensionless compressibility of $\kappa^{-1}=15.9835$ for water, Eq. \ref{eq:inverse_kappa} provides the key GCMe parametrization relationship
\begin{equation} \label{eq:gcm}
  A_{ij}=\frac{(15.9835N_\mathrm{m}-1)k_\mathrm{B}T}{0.998\rho},
\end{equation}
with $\rho$ and $N_\mathrm{m}$ being free parameters.

From a computational perspective, it is favorable to choose the lowest $\rho$ value that still satisfies the scaling relation since the number of pair interactions scales linearly with $\rho$ and the complexity of a single step of a MD simulation increases with the square of $\rho$. Fig. \ref{fig:water_p_excess_vs_rho}b indicates that $\rho=2.5d^{-3}$, which is indicated by the black vertical dotted line, is a reasonable choice. Similarly, for the coarse-graining level $N_\mathrm{m}$, we elected to apply the commonly used four-to-one mapping scheme where each simulation particle roughly encompasses the volume of a (CH\textsubscript{2})\textsubscript{3} group in a straight-chain alkane.\cite{groot_mesoscopic_2001,groot_electrostatic_2003,li_dissipative_2016} This real-space renormalization not only affects $A_{ij}$ via Eq. \ref{eq:gcm} and the length, mass, and time scales, which are now $d=(\rho d^3N_\mathrm{m})^{1/3}\times(0.275$ $\mathrm{nm})\approx0.592$ $\mathrm{nm}$, $m=N_\mathrm{m}\times(18.02$ $\mathrm{g/mol})\approx72.1$ $\mathrm{g/mol}$, and $\tau\approx3.18$ $\mathrm{ps}$, but also speeds up the simulation for a given system volume because there are now fewer particle positions to update.

Finally, to extend GCMe from melts to polymeric mixtures and solutions or to study liquid--liquid interfaces, a connection between GCMe and the Flory--Huggins theory\cite{flory_thermodynamics_1941,huggins_solutions_1941} can be made to reinterpret $A_{ij}$ in terms of an energetic $\chi$-parameter. This mapping is available in the SI.

\subsection*{Method of image charges}
In systems with charged species and parallel planar electrodes, such as electric double-layer capacitors (EDLCs), an important consideration is the proper treatment of the ion--electrode interactions to capture the correct electrostatic behavior at the interface. Historically, the two most frequently used methods were to enforce constant charges on the particles consitituting the electrodes, or allow those charges to fluctuate based on the charge distributions in their local environment via a variational procedure.\cite{siepmann_influence_1995} However, the former approach does not reproduce the electric double-layer (EDL) structures and dynamics obtained from simulations where the electrodes can polarize in response to nearby charge fluctuations,\cite{merlet_simulating_2013} and the latter is computationally expensive.

An alternative and efficient approach is the method of image charges, which accounts for the dielectric mismatch between the electrolyte and the bounding medium using image charge interactions.\cite{hautman_molecular_1989} An ion feels repulsion from or attraction to the interface when the electrode material has a lower or higher relative permittivity $\varepsilon_\mathrm{r,\,electrode}$, respectively, because the corresponding image charge on the other side of the interface carries a charge of
\begin{equation}
  q_{i,\,\mathrm{IC}}=z_ie\frac{\varepsilon_\mathrm{r}-\varepsilon_\mathrm{r,\,electrode}}{\varepsilon_\mathrm{r}+\varepsilon_\mathrm{r,\,electrode}},
\end{equation}
which has the same sign as the ion when $\varepsilon_\mathrm{r}>\varepsilon_\mathrm{r,\,electrode}$ and the opposite sign when $\varepsilon_\mathrm{r}<\varepsilon_\mathrm{r,\,electrode}$. In this work, we focus on the limiting cases of perfectly-conducting metal electrodes ($\varepsilon_\mathrm{r,\,electrode}=\infty$), where the image charges have equal but opposite charges to the ions they mirror, and nonmetal electrodes with the same permittivity as the electrolyte, where the image charges vanish.

\begin{figure}[ht]
\centering
  \includegraphics[width=3.25in]{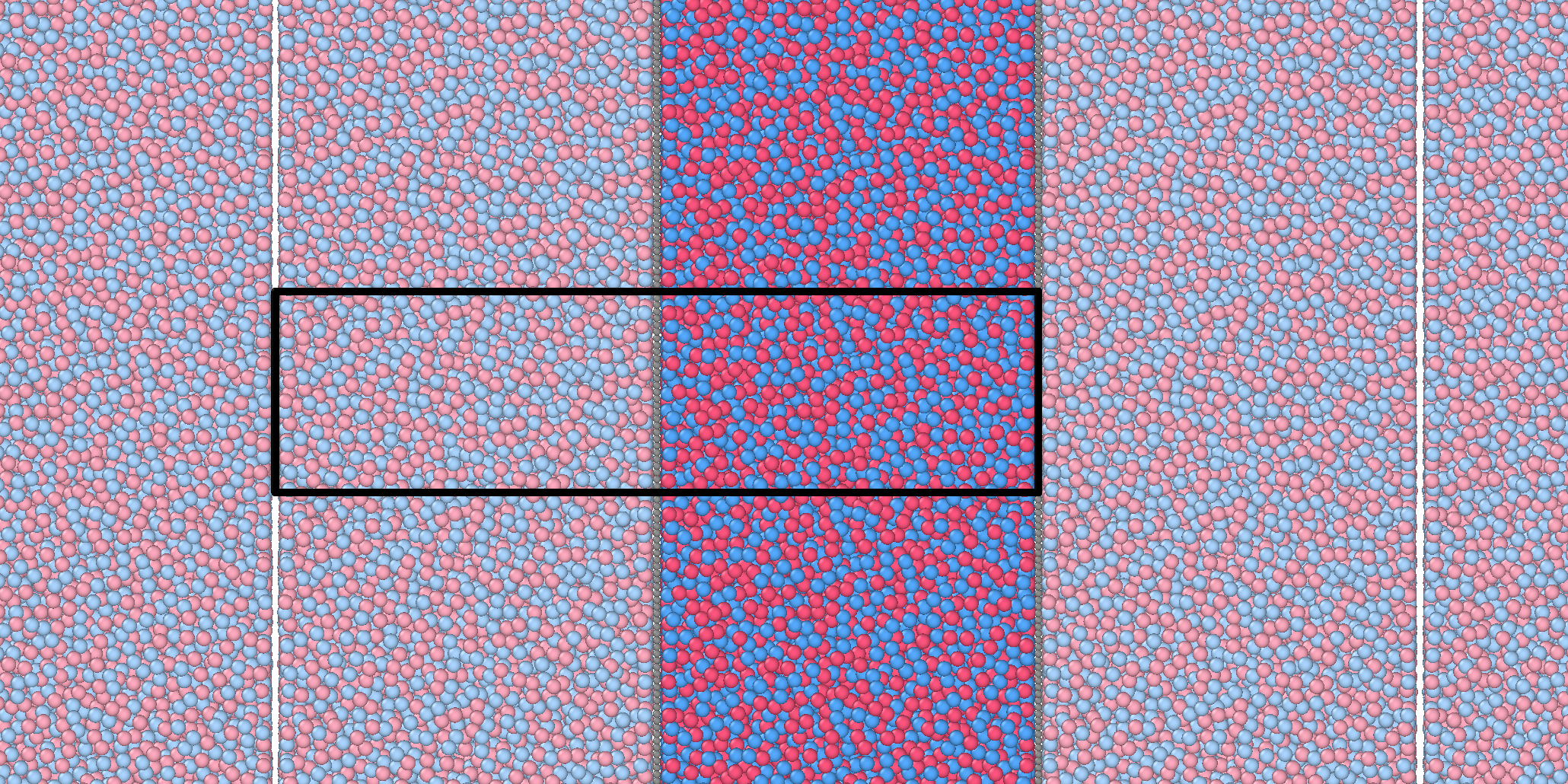}
  \caption{Two-dimensional representation of the three-dimensional periodic system. The unit cell containing the real and image systems is enclosed by the solid box. The blue and red colors represent the negative and positive charges, while the dark and light shades specify the real and image charges, respectively.}
  \label{fig:image_charges}
\end{figure}

The method of image charges has recently been implemented in OpenMM,\cite{son_image-charge_2021} fully leveraging its GPU acceleration to achieve performance that is orders of magnitude higher than that of traditional multithreaded CPU implementations.\cite{dwelle_constant_2019} In the simulation box of dimensions $L_x\times L_y\times2L_z$, there are two subsystems: the real system ($0<z<L_z$) that contains the ions and an image system ($-L_z<z<0$) that contains the image charges mirroring the ions. The basic idea of the method of image charges is that the ions in the real system are reflected across the electrode at $z=0$ and given opposite charges to create the corresponding image charges in the image system. As such, the overall system is electroneutral and has no net potential difference between $z=-L_z$ and $z=L_z$. With periodic boundary conditions, the subsystems form a unit cell that is repeated ad infinitum in all directions, as shown in Fig. \ref{fig:image_charges}. The overall system is periodic in the $x$- and $y$-directions to simulate a slab and in the $z$-direction since both electrodes are perfectly conducting. In each timestep, because the motion of the real ions and the image charges are coupled, the positions of the real ions are updated first by evaluating the forces acting on them from their interactions with each other and the image charges, and integrating the Langevin equations of motion. Then, the positions of the image charges are updated to reflect the new positions of the real ions.

\subsection*{Performance}
We have implemented GCMe in OpenMM,\cite{eastman_openmm_2017} a high-performance MD simulation toolkit, to take advantage of its class-leading GPU acceleration among the popular MD packages\cite{kondratyuk_gpu-accelerated_2021,wyrzykowski_gpu-based_2023} and leverage its extensibility and modularity to implement Eqs. \ref{eq:u_ex} and \ref{eq:u_elec} as custom pair potentials (available in our open-source MDCraft Python package).

To highlight the performance (or simulation throughput, defined as simulation time per elapsed real time) of our GCMe implementation, we benchmarked systems using the Weeks--Chandler--Andersen (WCA) and Coulomb potentials and GCMe with various boundary conditions and integration step sizes in both OpenMM 7.7.0 and LAMMPS (21 Nov 2023). LAMMPS\cite{thompson_lammps_2022} currently contains the only other GCMe implementation through a combination of \verb|pair_style gauss| in the base package and \verb|pair_style buck6d/coul/gauss/long| in the MOF-FF package.\cite{bureekaew_mofff_2013} For the method of image charges, we employed the GPU-only \verb|openmm_constV| OpenMM plugin\cite{son_image-charge_2021} and the CPU-only LAMMPS \verb|fix imagecharges| command.\cite{dwelle_constant_2019} All benchmarks were performed using a desktop computer with an Intel Core i9-10900K CPU and an NVIDIA RTX 3080 (10 GB) GPU running Ubuntu 20.04 LTS through Windows Subsystem for Linux. The results are tabulated in Table \ref{tbl:performance}.  

\begin{table*}[ht]
\caption{Comparison of simulation timesteps per second across systems with $N=1,000$ particles and varying reduced number densities $\rho^*$, models, boundary conditions, reduced step sizes $\tau^*$, and simulation toolkits.}
\label{tbl:performance}
\begin{tabular}{cllclr}
\toprule
$\rho^*$ & Model & System & $\tau^*$ & Toolkit & Timestep/s \\
\midrule
0.8 & WCA/Coulomb & Slab w/ image charges\textsuperscript{\emph{a}} & 0.005 & LAMMPS (OpenMP)\textsuperscript{\emph{d}} & 429 \\
 & & & & OpenMM (CUDA)\textsuperscript{\emph{e}} & 3,257 \\
\midrule
2.5 & GCMe & Bulk\textsuperscript{\emph{b}} & 0.005 & LAMMPS (OpenMP)\textsuperscript{\emph{f}} & 342 \\
 & & & & OpenMM (CUDA)\textsuperscript{\emph{e}} & 4,299 \\
 & & & 0.020 & LAMMPS (OpenMP)\textsuperscript{\emph{f}} & 320 \\
 & & & & OpenMM (CUDA)\textsuperscript{\emph{e}} & 4,250 \\
 & & Slab\textsuperscript{\emph{b},\emph{c}} & 0.005 & LAMMPS (OpenMP)\textsuperscript{\emph{f}} & 263 \\
 & & & & OpenMM (CUDA)\textsuperscript{\emph{e}} & 2,760 \\
 & & & 0.020 & LAMMPS (OpenMP)\textsuperscript{\emph{f}} & 255 \\
 & & & & OpenMM (CUDA)\textsuperscript{\emph{e}} & 2,832 \\
 & & Slab w/ image charges\textsuperscript{\emph{b},\emph{c}} & 0.005 & LAMMPS (OpenMP)\textsuperscript{\emph{f}} & 163 \\
 & & & & OpenMM (CUDA)\textsuperscript{\emph{e}} & 3,772 \\
 & & & 0.020 & LAMMPS (OpenMP)\textsuperscript{\emph{f}} & 152 \\
 & & & & OpenMM (CUDA)\textsuperscript{\emph{e}} & 3,689 \\
\bottomrule
\end{tabular}
\begin{tablenotes}[flushleft]
\item \textsuperscript{\emph{a}} The simulation system has dimensions of $8.0d\times7.8d\times20.0d$, where $d$ is the particle diameter.
\item \textsuperscript{\emph{b}} The simulation system has dimensions of $5.5d\times5.2d\times14.0d$.
\item \textsuperscript{\emph{c}} Each electrode contains 288 particles with size $d/2$ arranged in a hexagonal close-packed (HCP) lattice carrying charges such that the constant surface charge density is $\sigma_q=0.005\,\mathrm{e/nm}^2$. 
\item \textsuperscript{\emph{d}} The simulation was allocated two CPU threads (higher values led to simulation instability).
\item \textsuperscript{\emph{e}} The simulations were allocated a single GPU.
\item \textsuperscript{\emph{f}} The simulations were allocated eight CPU threads (higher values did not improve performance).
\end{tablenotes}
\end{table*}

For identical GCMe systems with different boundary conditions, our testing shows that our GPU-accelerated implementation in OpenMM is on average over an order of magnitude faster than the existing multithreaded LAMMPS implementation, with the largest speedup of $23\times$ observed in the slab system with image charges. When compared to systems utilizing the LJ-based Weeks--Chandler--Andersen (WCA) and point-charge Coulomb potentials, our GCMe implementation in OpenMM provides a $8\times$ performance uplift over the fastest OpenMP-accelerated LAMMPS method we had access to in our previous study.\cite{ye_coarse-grained_2022} Furthermore, larger step sizes can be taken in GCMe while still properly maintaining temperature control since the potential energy does not diverge when particles come into contact, unlike the WCA and point-charge Coulomb potentials. As such, we can achieve a further four-fold performance boost by using a step size of $\tau^*=0.02$ in GCMe, enabling an OpenMM simulation of a slab system with image charges that is up to $34\times$ faster than a comparable system simulated with WCA and Coulomb potentials. Remarkably, this speedup does not even account for the multiple orders of magnitude faster dynamics from the removal of the caging effect,\cite{groot_mesoscopic_2001} which further extends the physical time and length scales attainable with a soft-core model like GCMe.

\section*{Illustrative example}
\subsection*{Image charge effects on polyelectrolyte adsorption}
EDLCs have received significant interest in the recent years owing to their substantial power densities, fast charge and discharge rates, and sustainable cyclability compared to other energy storage devices, such as batteries and fuel cells.\cite{wang_latest_2017} The performance of an EDLC is strongly correlated with the EDL structure at the ion--electrode interface, which is largely dependent on the choice of the electrolyte and the electrode. The maximum energy $U$ an EDLC can store scales as $CV^2/2$, where $C$ is the capacitance and $V$ is the voltage, so electrolytes with large operational voltage windows, great mechanical stability, and high conductivity have been sought after. As a result, polyelectrolytes and polymerized ionic liquids have recently attracted considerable attention since they satisfy the design criteria with their wide electrochemical windows, outstanding thermal stability, and notable mechanical strength and conductivity.\cite{zhong_review_2015,zhang_mechanisms_2019} Indeed, experimental studies and computer simulations\cite{matsumoto_exceptionally_2017,lian_electrochemical_2018,eyvazi_molecular_2022} have shown that polyelectrolytes have exceptionally high EDL capacitance, making them prime electrolyte candidates for the next generation of EDLCs.

On the other hand, limited attention has been given to how the polarizability of the electrode material affects the adsorption behavior of the charged species in polyelectrolyte EDLCs, in part due to the technical challenges of efficiently conducting particle-based simulations while accounting for image charge effects. Existing studies have largely focused on how repulsive image charges and finite potential differences influence the polyelectrolyte adsorption behavior and energy storage. For example, Monte Carlo simulations by Wang et al.\cite{wang_adsorption_2023} have elucidated how repulsive image charges can either enhance or penalize polyelectrolyte adsorption depending on the polyelectrolyte charge fraction and surface charge distribution. MD efforts by Bagchi et al.\cite{bagchi_surface_2020} have revealed that repulsive image charges give rise to enhanced energy storage in polyelectrolyte EDLCs at low surface charge densities due to charge amplification. However, the effect of attractive image charges on the polyelectrolyte adsorption behavior is not well understood.

To this end, we use GCMe to study the adsorption behavior of polyanions on neutral parallel planar perfectly conducting and nonmetal electrodes in polyelectrolyte EDLCs. Simulations were carried out in the canonical ensemble with $N=96,000$ equisized particles and a temperature of $T=300$ $\mathrm{K}$. The salt-free electrolyte consists of polyanions with chain length $N_\mathrm{p}=60$ and their counterions, each with ion fraction $x_\mathrm{p}$, and solvent particles to fill the empty space. The chain connectivity in the polyanions is modeled by a harmonic bond potential
\begin{equation}
  u_\mathrm{harm}(r_{ij})=\frac{1}{2}k(x-b)^2,
\end{equation}
where $k=100(N_\mathrm{A}k_\mathrm{B}T)/d^2$ is the force constant and $b=0.8d$ is the equilibrium bond length, which coincides with where the radial distribution function for a neutral GCMe fluid (Fig. S2 in the SI) has its maximum. The solvent polarization is implicitly accounted for using a dielectric continuum with the relative permittivity of water, $\varepsilon_\mathrm{r}=78$. 

The real systems are initialized by first determining the $L_x$ and $L_y$ values that will accommodate the periodic placement of the electrode particles, which are half the size of the electrolyte particles and arranged in a hexagonal close-packed lattice, and then $L_z$, which is constrained by the number of particles and number density. For the parameters we chose, our real systems have dimensions of approximately $25d\times25.11d\times61.16d$. Then, the system is randomly filled with the polyanions, counterions, and solvent particles, and undergoes a local energy minimization. Finally, for the systems with perfectly conducting electrodes, the simulation box is doubled in width so that the real system can be reflected over the electrode at $z=0$ to generate the image charges. For the systems with nonmetal electrodes with no image charges, the $z$-dimension of the simulation box is instead tripled to introduce a sufficiently large void between periodic replicas in the $z$-direction so that the Yeh--Berkovitz correction\cite{yeh_ewald_1999,ballenegger_simulations_2009} can be used to remove the long-ranged electrostatic slab--slab interactions. In either case, because the simulations still have periodic boundary conditions, we can continue to use the fast PME method to evaluate the long-range electrostatic interactions, despite the lack of actual periodicity in the $z$-direction. 

For each simulation, system equilibration was performed over $5\times10^6$ timesteps of step size $t=0.02\tau$, and data were collected over at least an additional $2\times10^7$ timesteps, which is on the order of $1$ $\mu\mathrm{s}$.

\begin{figure}[!ht]
\centering
  \includegraphics[width=3.25in]{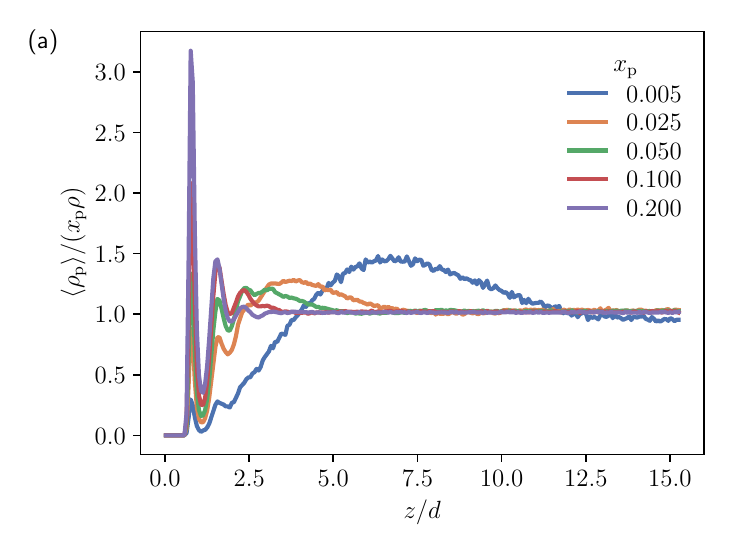}
  \includegraphics[width=3.25in]{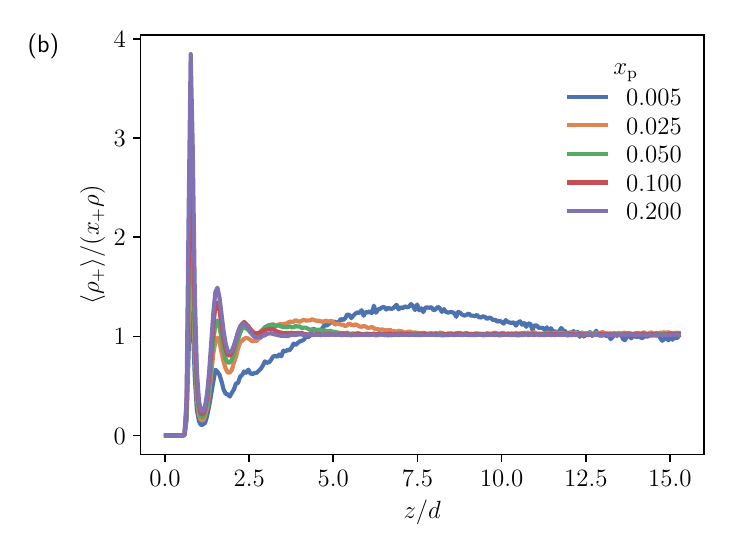}
  \includegraphics[width=3.25in]{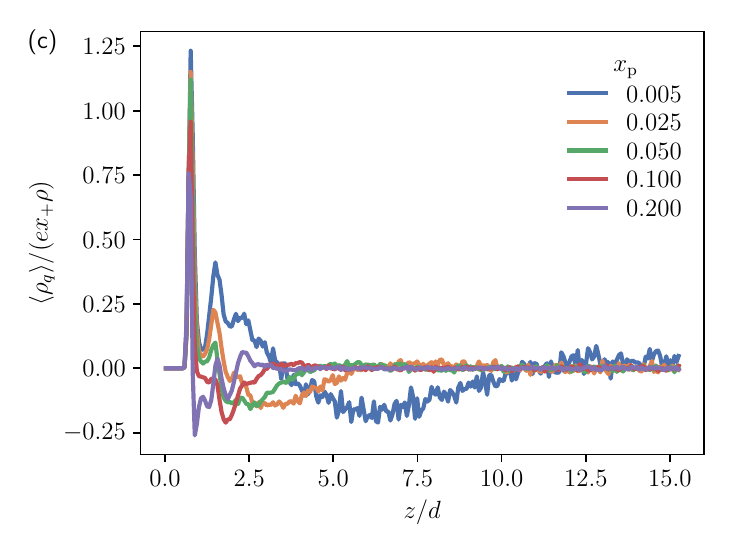}
  \caption{Ensemble-averaged (a) polyanion and (b) counterion number density profiles normalized by their respective total number densities, and (c) the charge density profile normalized by the total counterion charge density near an electrode for a salt-free system confined by nonmetal electrodes. The total number density differs from the bulk number density in the center of the system by less than 2\%.}
  \label{fig:pcs_slab}
\end{figure}

\begin{figure}[ht]
\centering
  \includegraphics[width=3.25in]{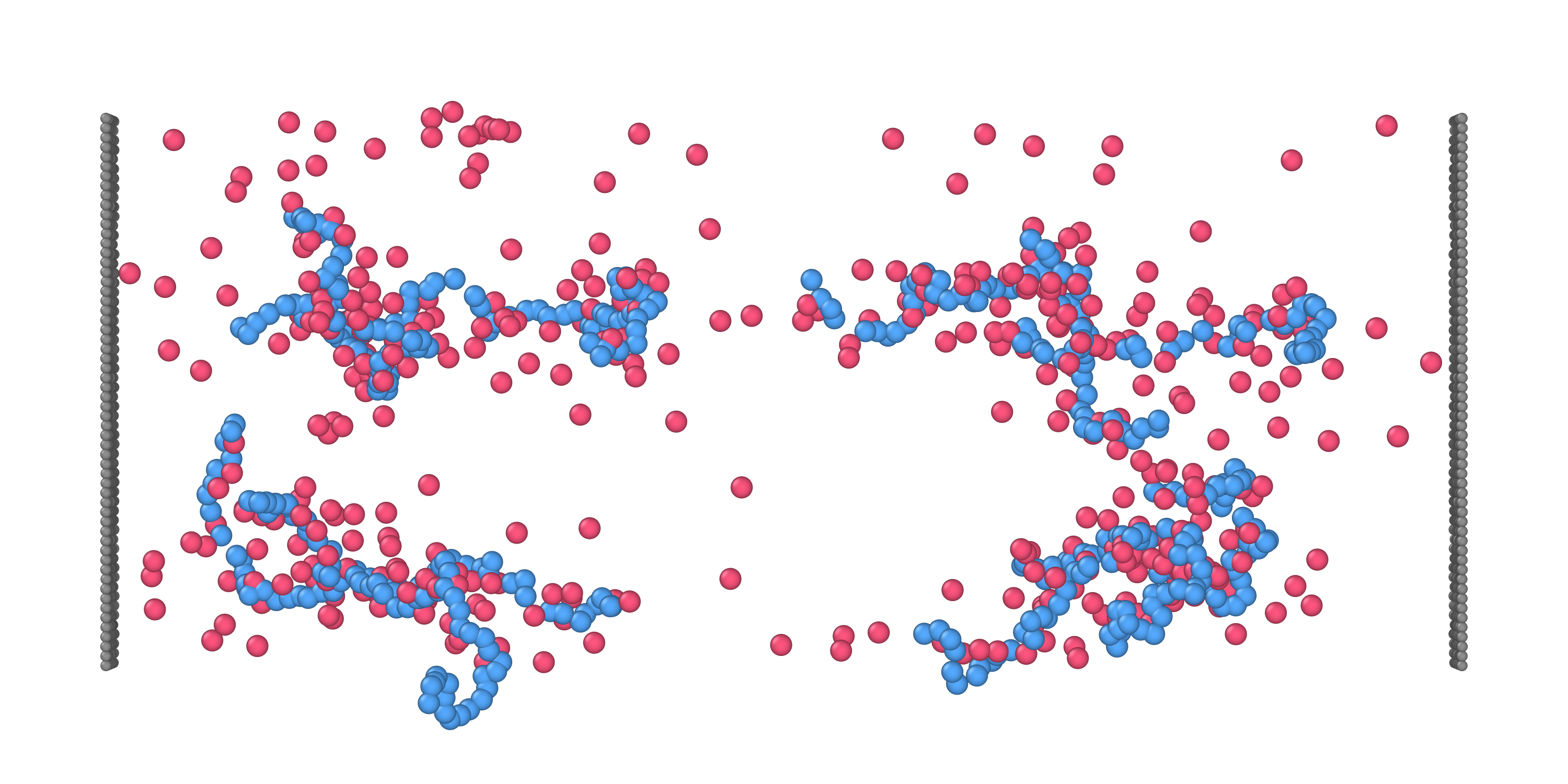}
  \caption{Representative simulation snapshot of the system with nonmetal electrodes at $x_\mathrm{p}=0.005$. Solvent particles are not shown.}
  \label{fig:slab_0.005}
\end{figure}

The polyanion and counterion number density profiles for the systems with nonmetal electrodes are presented in Fig. \ref{fig:pcs_slab}. As expected, they exhibit the same general trends as the density profiles obtained using classical polymer density functional theory (PDFT).\cite{jiang_density_2018,jiang_origin_2019} In systems with low polyanion fractions, such as the $x_\mathrm{p}=0.005$ system shown in Fig. \ref{fig:slab_0.005}, there is a net depletion of polyanions near the electrode due to the conformational entropy penalty for them to adsorb onto the surface. Instead, the polyanions aggregate a short distance away from the electrodes, as shown by the peak centered at $z = 7.5d$ in Fig. \ref{fig:pcs_slab}a. The counterions are also depleted in the vicinity of the electrode due to their favorable interactions with the polyanions, although to a lesser extent since they are nonbonded and do not suffer from an entropic penalty when they absorb onto the electrode. This local charge imbalance gives rise to an effective EDL, shown in Fig. \ref{fig:pcs_slab}c, where the positively-charged inner layers are screened by the negatively-charged outer layers. 

As $x_\mathrm{p}$ increases, the nominal Debye length $\lambda_\mathrm{D}=(4\pi\lambda_\mathrm{B}\rho\sum_{j=1}^Nx_j)^{-1/2}$, which is a measure of the range of an ion's electrostatic effects, decreases. This increased screening renders the repulsive polyanion--polyanion and counterion--counterion electrostatic interactions weaker, enabling both charged species to accumulate in greater numbers near the electrodes. As a result, there are strong peaks immediately next to, and oscillatory behavior away from, the electrode in our simulation density profiles, unlike those predicted by PDFT.\cite{jiang_origin_2019} Notably, GCMe is able to capture the accumulation of polyanions and counterions on the surfaces driven by the packing effect from the solvent particles, an effect not as pronounced in PDFT due to its treatment of the solvent using a local incompressibility condition.

\begin{figure}[h!]
\centering
  \includegraphics[width=3.25in]{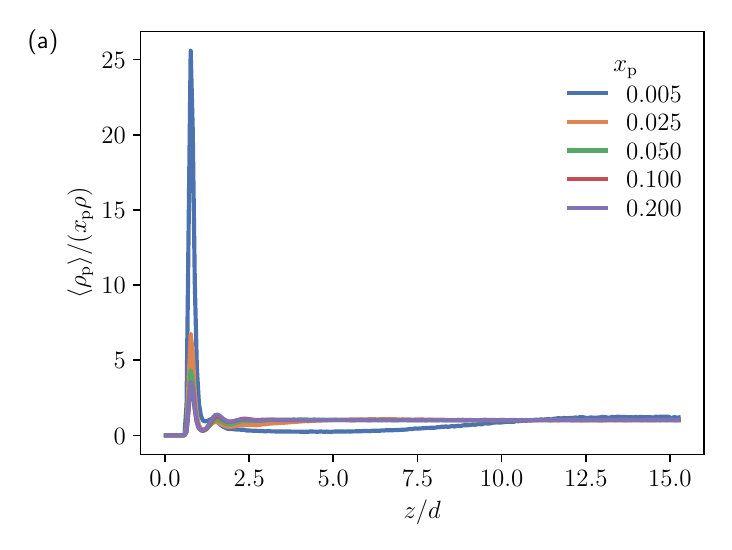}
  \includegraphics[width=3.25in]{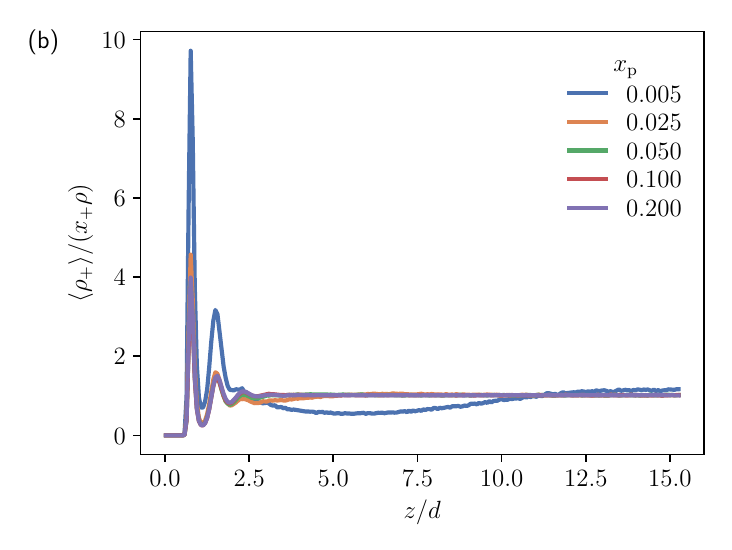}
  \includegraphics[width=3.25in]{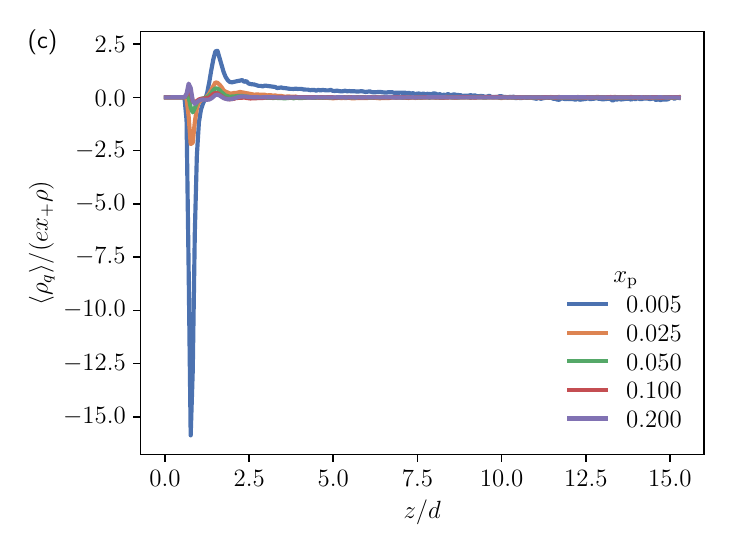}
  \caption{Ensemble-averaged (a) polyanion and (b) counterion number density profiles normalized by their respective total number densities, and (c) the charge density profile normalized by the total counterion charge density near an electrode for a salt-free system confined by perfectly conducting electrodes. The total number density differs from the bulk number density in the center of system by less than 2\%.}
  \label{fig:pcs_ic}
\end{figure}

\begin{figure}[ht]
\centering
  \includegraphics[width=3.25in]{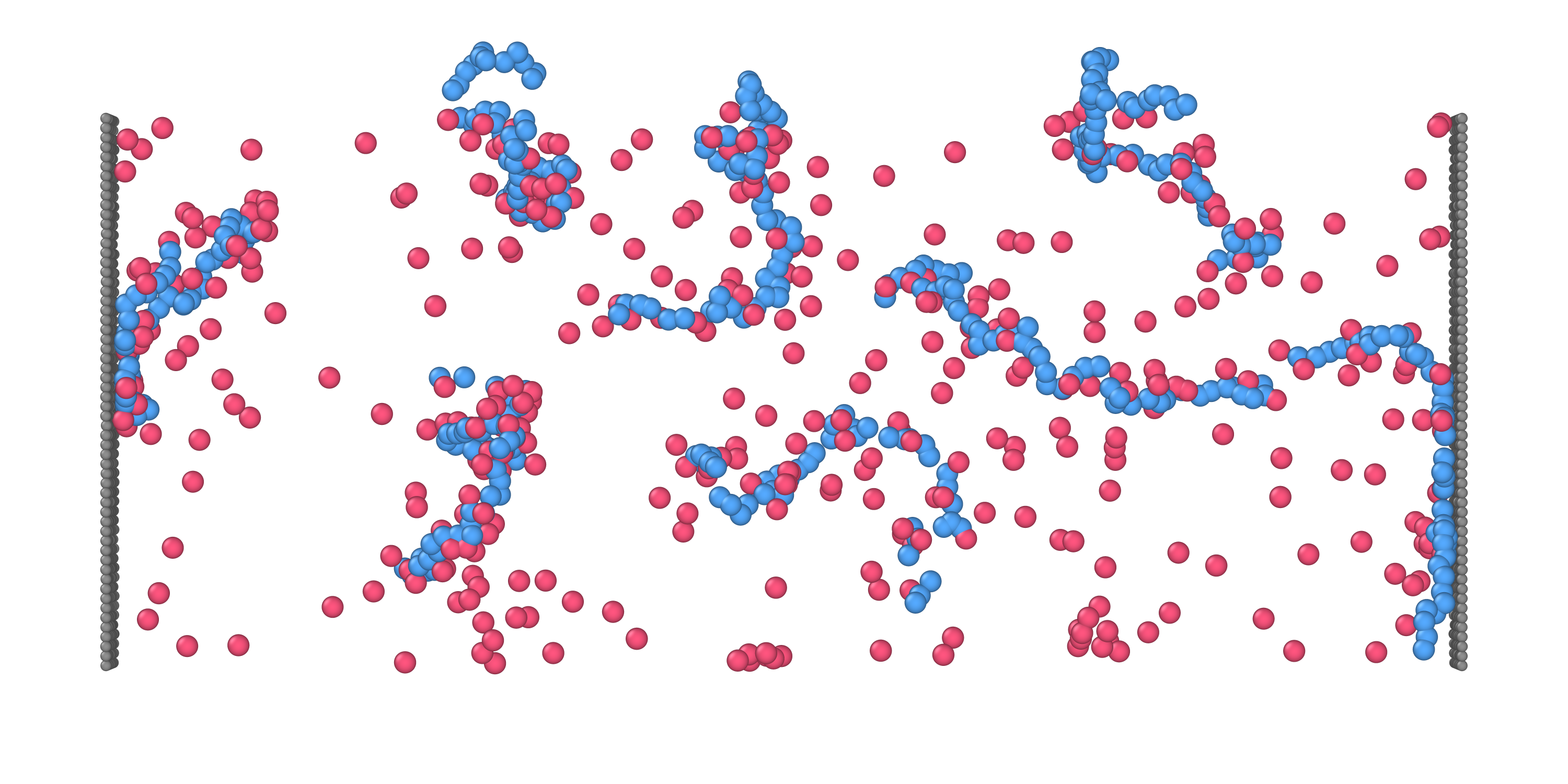}
  \caption{Representative simulation snapshot of the real system with perfectly conducting electrodes at $x_\mathrm{p}=0.005$. Solvent particles are not shown.}
  \label{fig:ic_0.005}
\end{figure}

With the inclusion of attractive image charges, the arrangement of polyanions and counterions near the perfectly conducting electrodes is vastly different. In what appears to be a complete reversal from the systems with nonmetal electrodes, there is strong accumulation of both species on the electrode and, consequently, a depletion in the outer layers of the EDL. At the low $x_\mathrm{p}=0.005$, the singular intial peak in Fig. \ref{fig:pcs_ic}a and the simulation snapshot in Fig. \ref{fig:ic_0.005} clearly show that the polyanions prefer to be adsorbed on the electrode, suggesting that the attractive image charge interactions can overcome the entropic penalty. Fig. \ref{fig:pcs_ic}b shows that there is also an influx of counterions due to the attractive electrostatic polyanion--counterion interactions, but not in equal proportion to the polyanions. As a result, there is now a considerable net negative charge next to the electrode, as shown in Fig. \ref{fig:pcs_ic}c, and a second diffuse layer of counterions is needed to neutralize the first polyanion layer. Additionally, Fig. S6 in the SI shows that systems with low ion fractions now have positive bulk electrostatic potentials, as opposed to the negative ones observed in comparable systems with nonmetal electrodes.

\begin{figure}[ht]
\centering
  \includegraphics[width=3.25in]{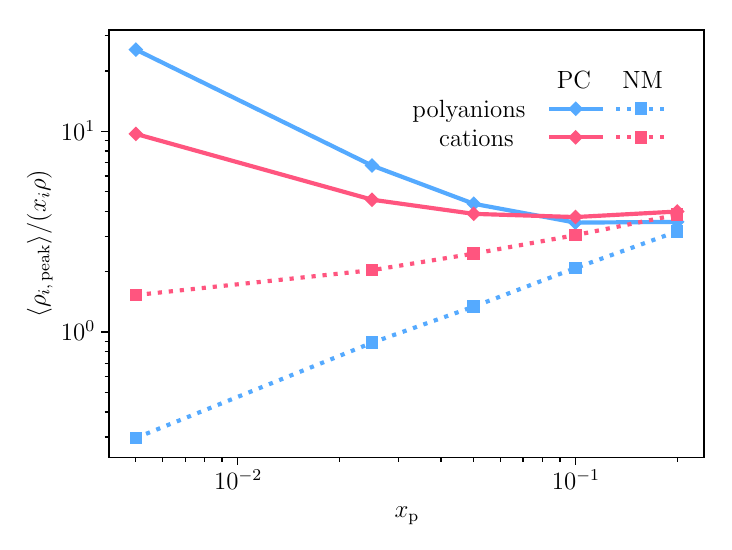}
  \caption{Ensemble-averaged polyanion and counterion number densities $\rho_{i,\,\mathrm{peak}}$ in the first adsorption layer next to the electrode as functions of the polyanion fraction for perfectly conducting (PC) and nonmetal (NM) boundaries.}
  \label{fig:pcs_peaks}
\end{figure}

Perhaps surprisingly, when $x_\mathrm{p}$ first begins to increase, the number of polyanions and cations adsorbed on the electrode relative to the bulk instead decreases. A simple explanation is that the increased screening length diminishes the image charge interactions, lessening the driving force for the adsorption of either charged species to the electrodes. This effect is so much more significant than the coinciding weakening of the repulsive electrostatic interactions between like species that the net result is a substantial decrease in the amount of both polyanions and counterions in the EDL. As the polyanion fraction increases further, i.e., $x_\mathrm{p}\geq0.05$, the changes in the polyanion and counterion distributions near and away from the electrode become less pronounced, as evidenced by the converging number densities in Fig. \ref{fig:pcs_ic}. Particularly, the systems with perfectly conducting electrodes end up having a comparable makeup as the systems with nonmetal electrodes at high $x_\mathrm{p}$. This is best illustrated in Fig. \ref{fig:pcs_peaks}, where the polyanion and counterion number densities in the first layer of the EDL for the nonmetal and systems with perfectly conducting electrodes reach similar values at $x_\mathrm{p}=0.2$.

Therefore, the attractive image charge interactions from perfectly conducting electrodes can have a marked effect on the EDL structure in polyelectrolyte EDLCs, especially at low ion fractions. While the systems with nonmetal electrodes have a moderate net positive layer next to the electrode due to an uneven depletion of polyanions and counterions, the systems with perfectly conducting electrodes have a sizeable net negative adsorption layer due to the enhanced adsorption of polyanions driven by their net electrostatic attraction to the electrode by the image charges.

\section*{Conclusions}
In conclusion, we have developed and implemented GCMe, a soft-core model for MD simulations that combines the GCM and smeared electrostatic interactions and is directly compatible with the models used in recent theoretical studies\cite{guenza_cooperative_2001,guenza_intermolecular_2002,wang_fluctuation_2010,villet_efficient_2014,delaney_recent_2016,mccarty_complete_2019,shen_learning_2020} to study charged soft matter systems. The key feature of GCMe lies in its soft interaction potentials, which have clear physical origins and can accurately capture the weaker packing effects and short-range electrostatics in deformable organic molecules and systems with strong coarse-graining. Moreover, the excluded volume and electrostatic potentials do not diverge when particles overlap, allowing larger simulation timesteps to be taken. This, when combined with the optimized GPU-accelerated framework in the high-performance OpenMM toolkit, gives our GCMe implementation class-leading efficiency among similar coarse-grained simulation methods. Our benchmarks show that GCMe in OpenMM is at least $8\times$ faster than the only existing GCMe implementation currently available, and can be up to $34\times$ faster than the coarse-grained methods we had access to in our recent study.\cite{ye_coarse-grained_2022} This significant performance improvement enables simulating systems with hundreds of thousands of charged particles (corresponding to millions of atoms) over time scales on the order of milliseconds in just a few hours on modern computer hardware.

To test and demonstrate the capabilities of GCMe, we examined systems consisting of polyanions, their counterions, and solvent particles confined between uncharged planar parallel nonmetal or perfectly conducting metal electrodes, with the electrode polarizability accounted for by the method of image charges. For systems with nonmetal electrodes, we were able to qualitatively reproduce the depletion of polyanions and counterions from the surfaces that the PDFT\cite{jiang_origin_2019} predicted. Notably, the GCMe simulations showed that there is a growing accumulation of both charged species on the surface as the ion fraction $x_\mathrm{p}$ increases, a feature missing in PDFT due to the density smearing in its incompressibility condition. For systems with perfectly conducting electrodes, which cannot be represented by a mean-field treatment, we found a complete reversal in the structure and charge of the effective EDLs. Peculiarly, there is a strong accumulation of both polyanions and counterions on the electrodes, suggesting that the attractive image charge interactions are favorable enough to overcome the conformational entropy penalty from the polyanion adsorption. As $x_\mathrm{p}$ increases and the nominal screening length decreases, the image charge interactions weaken and the EDL structure begin to resemble that from a high $x_\mathrm{p}$ system with nonmetal electrodes. Given the contrasting adsorption behavior in the systems with perfectly conducting electrodes, we anticipate very different EDL structures when the electrodes are charged. As such, we plan to quantify the energy storage and charging/discharging dynamics of these polyanion--counterion--solvent EDLCs in a future study.

We hope that the previous illustrative example has exemplified the applicability of GCMe. In principle, GCMe can be extended to more realistic systems by rigorously parametrizing it with important effects and properties accounted for, like particle size asymmetry, solvent type and quality, and explicit ion polarizability. Then, with the system-specific parametrization and its innate efficiency, we expect GCMe to be able to predict and explore rich bulk and interfacial phenomena in a wide variety of large soft matter systems over long time scales.

\begin{acknowledgement}
We thank Alejandro Gallegos, Dorian Bruch, and Pierre Walker for helpful discussions. This work is supported in part by funding from the Hong Kong Quantum AI Lab, AIR@InnoHK of the Hong Kong Government.
\end{acknowledgement}

\begin{suppinfo}
Mathematical derivations and supplemental figures are available in the SI.
\end{suppinfo}

\bibliography{bibliography}

\end{document}


\section*{Models and methods}
\subsection*{Derivation of the excluded volume interaction potential}
For a particle $i$ with mass $m_i$, its smeared density $\rho_i$ over radius $\sigma_i$ is
\begin{equation*}
  \rho_i(\mathbf{r})=\left(\frac{3}{2\pi\sigma_i^2}\right)^{3/2}\exp{\left[-\frac{3(\mathbf{r}-\mathbf{r}_i)^2}{2\sigma_i^2}\right]}
\end{equation*}
The excluded volume interaction potential between two particles $i$ and $j$ is
\begin{equation*}
  \begin{split}
    u_{ex}(r_{ij})&=A_{ij}\int\rho_i(\mathbf{r})\rho_j(\mathbf{r})\,d\mathbf{r}\\
    &=A_{ij}\left(\frac{3}{2\pi\sigma_i\sigma_j}\right)^3\int{\exp{\left[-\frac{3(\mathbf{r}-\mathbf{r}_i)^2}{2\sigma_i^2}\right]}\exp{\left[-\frac{3(\mathbf{r}-\mathbf{r}_j)^2}{2\sigma_j^2}\right]}}\,d\mathbf{r}
  \end{split}
\end{equation*}
Evaluating the Gaussian integral over three dimensions gives
\begin{equation*}
  \begin{split}
    u_\mathrm{ex}(r_{ij})=A_{ij}\left(\frac{3}{2\pi\sigma_{ij}^2}\right)^{3/2}\exp{\left[-\frac{3\left(\mathbf{r}_i-\mathbf{r}_j\right)^2}{2\sigma_{ij}^2}\right]}
  \end{split}
\end{equation*}
where $\sigma_{ij}^2=\sigma_i^2+\sigma_j^2$.

\subsection*{Derivation of the smeared electrostatic interaction potential}
The Fourier and inverse Fourier transforms are
\begin{equation*}
  \hat{f}(\mathbf{k})=\int{f(\mathbf{r})\exp{(-i\mathbf{k}\cdot\mathbf{r})}}\,d\mathbf{r}\quad\longleftrightarrow\quad f(\mathbf{r})=\frac{1}{(2\pi)^3}\int{\hat{f}(\mathbf{k})\exp{(i\mathbf{k}\cdot\mathbf{r})}}\,d\mathbf{k}
\end{equation*}
For an ion $i$ with charge $q_i=z_ie$, its smeared charge density $\rho_i$ over radius $a_i$ is
\begin{equation} \label{eq:elec_density}
  \rho_i(\mathbf{r})=\frac{z_ie}{(2a_i^2)^{3/2}}\exp{\left[-\frac{\pi\left(\mathbf{r}-\mathbf{r}_i\right)^2}{2a_i^2}\right]}
\end{equation}
The electrostatic interaction potential between two particles $i$ and $j$ is
\begin{equation} \label{eq:elec_pair}
  u_\mathrm{Coul}(r_{ij})=\int{\rho_i\left(\mathbf{r}\right)C\left(\mathbf{r}-\mathbf{r}^\prime\right)\rho_j\left(\mathbf{r}^\prime\right)}\,d\mathbf{r}^\prime\,d\mathbf{r}
\end{equation}
where $C(\mathbf{r}-\mathbf{r}^\prime)=(4\pi\varepsilon_0\varepsilon_\mathrm{r}|\mathbf{r}-\mathbf{r}^\prime|)^{-1}$ is the Coulomb operator. 

With $r=\left|\mathbf{r}\right|$ and the change of variable $\bar{\mathbf{r}}=\mathbf{r}-\mathbf{r}^\prime$, the Fourier transform of $C\left(\mathbf{r}-\mathbf{r}^\prime\right)$ multiplied by a Yukawa-type function $\exp{\left(-\lambda\left|\bar{\mathbf{r}}\right|\right)}$ that tends to $1$ as $\lambda\rightarrow 0$ is
\begin{equation} \label{eq:elec_coul_op_}
  \begin{split}
    \hat{C}(\mathbf{k})&=\lim_{\lambda\rightarrow0}{\int{\frac{1}{4\pi\varepsilon_0\varepsilon_\mathrm{r}\left|\bar{\mathbf{r}}\right|}\exp{\left(-i\mathbf{k}\cdot\bar{\mathbf{r}}\right)}\exp{\left(-\lambda\left|\bar{\mathbf{r}}\right|\right)}}\,d\bar{\mathbf{r}}}\\
    &=\frac{1}{4\pi\varepsilon_0\varepsilon_\mathrm{r}}\lim_{\lambda\rightarrow0}{\int\frac{\exp{\left(-ik\bar{r}\cos{\theta}-\lambda\bar{r}\right)}}{\bar{r}}\,d\bar{\mathbf{r}}}
  \end{split}
\end{equation}
Integrating Eq. \ref{eq:elec_coul_op_} in the spherical coordinate system using an $u$-substitution $u=\cos{\theta}$ and then by parts,
\begin{equation} \label{eq:elec_coul_op_fourier}
  \begin{split}
    \hat{C}(\mathbf{k})&=\frac{1}{4\pi\varepsilon_0\varepsilon_\mathrm{r}}\lim_{\lambda\rightarrow0}{\int_{0}^{\infty}{{\bar{r}}^2\int_{0}^{\pi}{\sin{\theta}\int_{0}^{2\pi}\frac{\exp{\left(-ik\bar{r}\cos{\theta}-\lambda\bar{r}\right)}}{\bar{r}}\,d\varphi}\,d\theta}\,d\bar{r}}\\
    &=\frac{1}{\varepsilon_0\varepsilon_\mathrm{r}k}\lim_{\lambda\rightarrow0}{\int_{0}^{\infty}{\sin{\left(k\bar{r}\right)}\exp{\left(-\lambda\bar{r}\right)}}\,d\bar{r}}\\
    &=\frac{1}{\varepsilon_0\varepsilon_\mathrm{r}k}\lim_{\lambda\rightarrow0}{\frac{k}{\lambda^2+k^2}}
  =\frac{1}{\varepsilon_0\varepsilon_\mathrm{r}k^2}
  \end{split}
\end{equation}
With the change of variable $\bar{\mathbf{r}}\equiv\mathbf{r}-\mathbf{r}_i$, the Fourier transform of Eq. \ref{eq:elec_density} is
\begin{equation} \label{eq:elec_density_fourier}
  {\hat{\rho}}_i(\mathbf{k})=\frac{z_ie}{(2a_i^2)^{3/2}}\int\exp{\left[-\left(\frac{\pi\bar{\mathbf{r}}\cdot\bar{\mathbf{r}}}{2a_i^2}+i\mathbf{k}\cdot\bar{\mathbf{r}}\right)\right]}\,d\bar{\mathbf{r}}
  =z_ie\exp{\left(-\frac{a_i^2k^2}{2\pi}\right)}
\end{equation}
Substituting Eqs. \ref{eq:elec_coul_op_fourier} and \ref{eq:elec_density_fourier} into Eq. \ref{eq:elec_pair} and taking the inverse Fourier transform by integrating over all $\mathbf{r}$ and $\mathbf{r}^\prime$ and using the definition of a delta function $\delta(\mathbf{r})=(2\pi)^{-3}\int\exp{(i\mathbf{k}\cdot\mathbf{r})}\,d\mathbf{k}$,
\begin{equation} \label{eq:elec_pair_}
  \begin{split}
  u_\mathrm{elec}(r_{ij})&=\frac{z_iz_je^2}{(2\pi)^9\varepsilon_0\varepsilon_\mathrm{r}}\int\frac{\mathrm{e}^{-(a_i^2k_1^2+a_j^2k_3^2)/(2\pi)}\mathrm{e}^{i\mathbf{k}_1\cdot\left(\mathbf{r}-\mathbf{r}_i\right)+i\mathbf{k}_2\cdot\left(\mathbf{r}-\mathbf{r}^\prime\right)+i\mathbf{k}_3\cdot\left(\mathbf{r}^\prime-\mathbf{r}_j\right)}}{k_2^2}\,d\mathbf{r}^\prime\,d\mathbf{r}\,d\mathbf{k}_1\,d\mathbf{k}_2\,d\mathbf{k}_3\\
  &=\frac{z_iz_je^2}{(2\pi)^3\varepsilon_0\varepsilon_\mathrm{r}}\int\frac{\exp{\left[ikr_{ij}\cos{\theta}-a_{ij}^2k^2/(2\pi)\right]}}{k^2}\,d\mathbf{k}
  \end{split}
\end{equation}
where $a_{ij}^2=a_i^2+a_j^2$ and $r_{ij}=\left|\mathbf{r}_i-\mathbf{r}_j\right|$.\\
Integrating Eq. \ref{eq:elec_pair_} in the spherical coordinate system using an $u$-substitution $u=\cos{\theta}$, the definition $2\sin{(kr_{ij})}/k=\int_{-r_{ij}}^{r_{ij}}\exp{(iks)}\,ds$, and a change of variables $s=(2^{1/2}a_{ij}/\pi^{1/2})t$,
\begin{equation*} \label{eq:elec_s22}
  \begin{split}
    u_\mathrm{elec}(r_{ij}) &=\frac{z_iz_je^2}{(2\pi)^3\varepsilon_0\varepsilon_\mathrm{r}}\int_{0}^{\infty}{k^2\int_{0}^{\pi}{\sin{\theta}\int_{0}^{2\pi}\frac{\exp{\left[ikr_{ij}\cos{\theta}-a_{ij}^2k^2/(2\pi)\right]}}{k^2}\,d\varphi}\,d\theta}\,dk\\
    &=\frac{z_iz_je^2}{(2\pi)^2\varepsilon_0\varepsilon_\mathrm{r}r_{ij}}\int_{0}^{\infty}\frac{2\sin{(kr_{ij})}}{k}\exp{\left(-\frac{a_{ij}^2k^2}{2\pi}\right)}\,dk\\
    &=\frac{z_iz_je^2}{2^{3/2}\pi\varepsilon_0\varepsilon_\mathrm{r}a_{ij}r_{ij}}\int_{0}^{r_{ij}}\exp{\left(-\frac{\pi}{2a_{ij}^2}s^2\right)}\,ds\\
    &=\frac{z_iz_je^2}{4\pi\varepsilon_0\varepsilon_\mathrm{r}r_{ij}}\erf\left(\frac{\pi^{1/2}}{2^{1/2}a_{ij}}r_{ij}\right)
  \end{split}
\end{equation*}

\subsection*{Limits}
When the separation distance between particles become very large ($r_{ij}\rightarrow\infty$) or when the electrostatic smearing radius approaches zero ($a_{ij}\rightarrow0$), the smeared electrostatic potential converges to the Coulomb potential as expected:
\begin{equation*}
  \lim_{r_{ij}\rightarrow\infty}u_\mathrm{elec}(r_{ij})=\frac{z_iz_je^2}{4\pi\varepsilon_0\varepsilon_\mathrm{r}r_{ij}}\lim_{r_{ij}\rightarrow\infty}\left[\erf\left(\frac{\pi^{1/2}}{2^{1/2}a_{ij}}r_{ij}\right)\right]=\frac{z_iz_je^2}{4\pi\varepsilon_0\varepsilon_\mathrm{r}r_{ij}}
\end{equation*}
\begin{equation*}
  \lim_{a_{ij}\rightarrow 0}u_\mathrm{elec}(r_{ij})=\frac{z_iz_je^2}{4\pi\varepsilon_0\varepsilon_\mathrm{r}r_{ij}}\lim_{a_{ij}\rightarrow0}\left[\erf\left(\frac{\pi^{1/2}}{2^{1/2}a_{ij}}r_{ij}\right)\right]=\frac{z_iz_je^2}{4\pi\varepsilon_0\varepsilon_\mathrm{r}r_{ij}}
\end{equation*}
When the separation distance between particles become very small ($r_{ij}\rightarrow0$), the excluded volume and smeared electrostatic potentials have the values
\begin{equation*}
  \lim_{r_{ij}\rightarrow 0}u_\mathrm{ex}(r_{ij})=A_{ij}\left(\frac{3}{2\pi\sigma_{ij}^2}\right)^{3/2}\lim_{r_{ij}\rightarrow 0}\left[\exp{\left(-\frac{3}{2\sigma_{ij}^2}r_{ij}^2\right)}\right]=A_{ij}\left(\frac{3}{2\pi\sigma_{ij}^2}\right)^{3/2}
\end{equation*}
\begin{equation*}
  \erf{\left(\frac{\pi^{1/2}}{2^{1/2}a_{ij}}r_{ij}\right)}=\frac{2^{1/2}}{a_{ij}}r_{ij}+\mathcal{O}(r_{ij}^3)\longrightarrow\lim_{r_{ij}\rightarrow0}u_\mathrm{elec}(r_{ij})\approx\frac{z_iz_je^2}{4\pi\varepsilon_0\varepsilon_\mathrm{r}r_{ij}}\left(\frac{2^{1/2}}{a_{ij}}r_{ij}\right)=\frac{z_iz_je^2}{2^{3/2}\pi\varepsilon_0\varepsilon_\mathrm{r}a_{ij}}
\end{equation*}
If ions $i$ and $j$ have the same Born radii $a=a_i=a_j$, the smeared electrostatic potential as $r_{ij}\rightarrow0$ becomes
\begin{equation*}
  \lim_{r_{ij}\rightarrow0}u_\mathrm{elec}(r_{ij})\approx\frac{z_iz_je^2}{4\pi\varepsilon_0\varepsilon_\mathrm{r}a}
\end{equation*}
The self-energy of an ion $i$ is
\begin{equation} \label{eq:self_energy}
  u_\mathrm{elec,\,self}=\frac{1}{2}\lim_{r_{ii}\rightarrow0}u_\mathrm{elec}(r_{ii})\approx\frac{z_i^2e^2}{8\pi\varepsilon_0\varepsilon_\mathrm{r}a_i}
\end{equation}
which is simply the Born energy.\cite{wang_fluctuation_2010}

\subsection*{Ewald summation}
\begin{figure}
\centering
  \begin{subfigure}{3.226in}
  \includegraphics[width=3.226in]{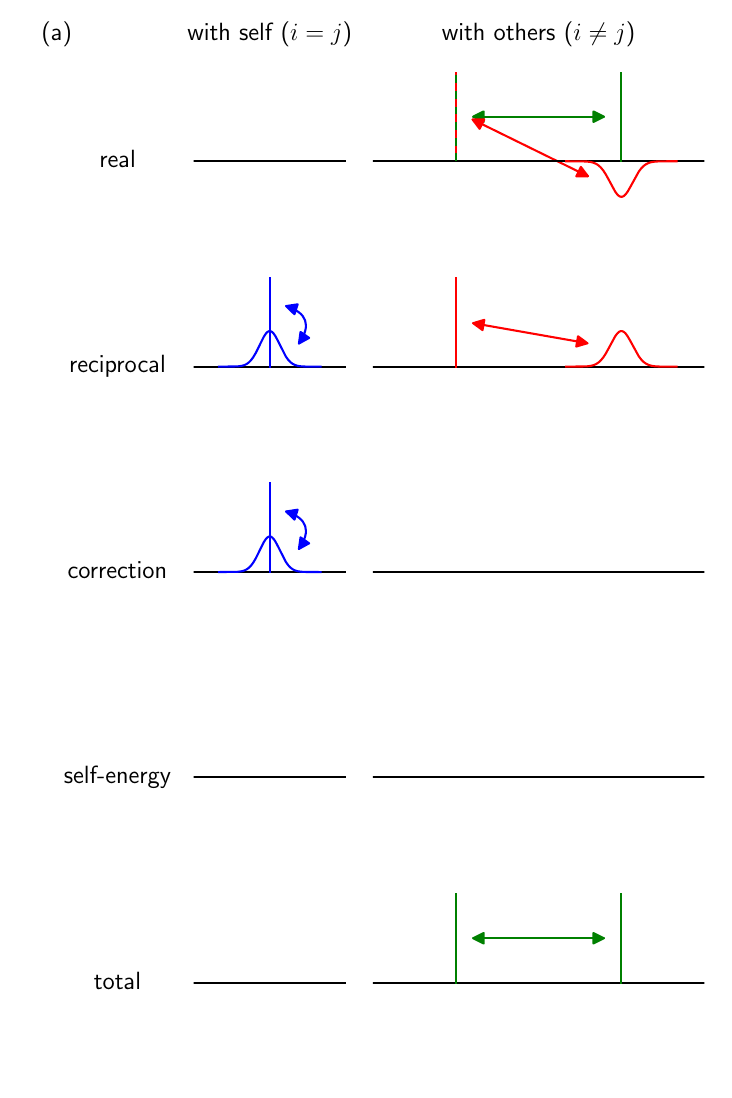}
  \end{subfigure}
  \begin{subfigure}{3.226in}
  \includegraphics[width=3.226in]{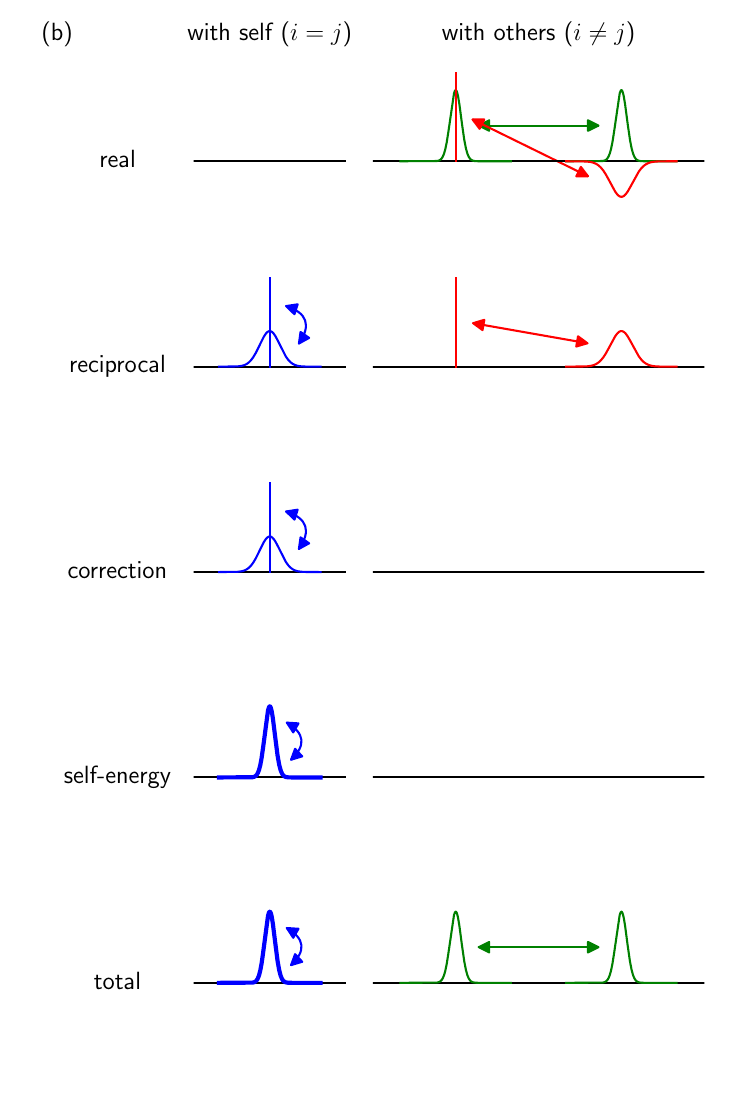}
  \end{subfigure}
  \caption{Schematic representations of the real-space, reciprocal-space, correction, and self-energy contributions to the Ewald summations for (a) point charges and (b) GCMe smeared charges. The vertical lines, narrow distributions, and wide distributions represent point charges, GCMe smeared charges, and screening/compensating charge distributions, respectively, each with charge magnitude $|q_i|=e$. Note that the correction term is subtracted in the summation.}
  \label{fig:si_ewald}
\end{figure}

\subsection*{Most probable pair separation distance}
\begin{figure}[H]
\centering
  \includegraphics[width=3.25in]{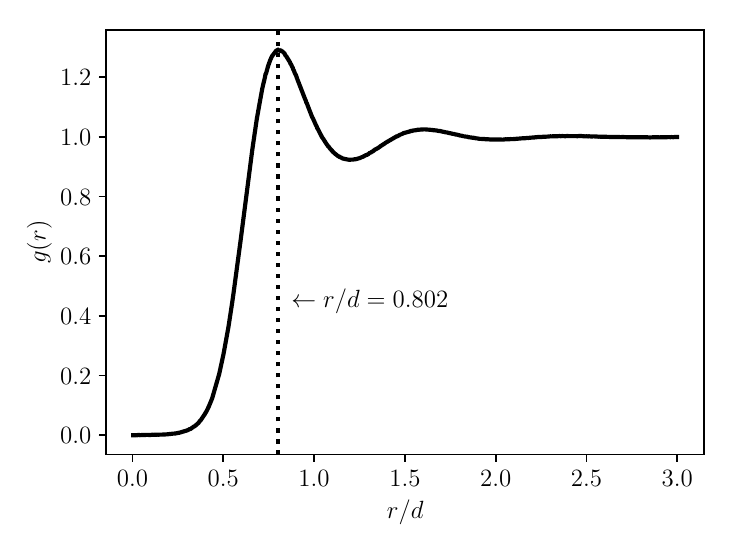}
  \caption{Radial distribution function $g(r)$ for uncharged nonbonded particles interacting via the parameterized GCMe with $N_\mathrm{m}=4$. The oscillations indicate a pronounced liquid structure. The first peak, indicated by the dotted line, is the most probable separation distance between two particles.}
  \label{fig:si_water_rdf}
\end{figure}

\subsection*{Flory--Huggins parameter mapping}
\begin{figure}[h]
\centering
  \includegraphics[width=3.25in]{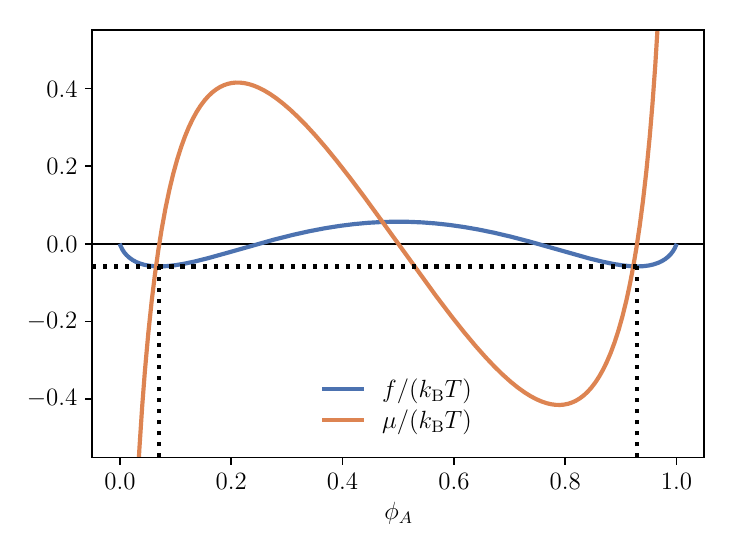}
  \caption{Free energy $f$ (solid line) and chemical potential $\mu$ (dashed line) in the FH theory for $N_\mathrm{p}=1$ and $\chi=3$. The dotted lines indicate the coexistence volume fractions and the corresponding free energy.}
  \label{fig:water_fh_theory}
\end{figure}

Since the GCM is phenomenologically equivalent to and can be considered a continuous version of the lattice-based Flory--Huggins (FH) theory,\cite{flory_thermodynamics_1941,huggins_solutions_1941} it can be used to study liquid--liquid and liquid--solid interfaces. For a binary polymer mixture of $A$ and $B$ chains, the FH theory gives a free energy per unit volume $f$ of
\begin{equation*}
  \frac{f}{k_\mathrm{B}T}=\frac{\phi_A}{N_{\mathrm{p},\,A}}\ln{\phi_A}+\frac{\phi_B}{N_{\mathrm{p},\,B}}\ln{\phi_B}+\chi\phi_A\phi_B,
\end{equation*}
where $\phi_A$ and $\phi_B$ are the volume fractions of components $A$ and $B$ (with $\phi_A+\phi_B=1$ such that all lattice sites are filled), $N_{\mathrm{p},\,A}$ and $N_{\mathrm{p},\,B}$ are the chain lengths of the $A$ and $B$ chains, and $\chi$ is a free mixing parameter. The sign of $\chi$ dictates the nature of the interactions between the $A$ and $B$ components: $A$--$A$ and $B$--$B$ contacts are favored when $\chi > 0$ while $A$--$B$ interactions are preferred when $\chi < 0$. When $\chi$ gets sufficiently large, there are two minima separated by a maximum in the free energy curve, as shown in Fig. \ref{fig:water_fh_theory}, indicating an equilibrium phase-separated state with $A$- and $B$-rich domains. If the chain lengths are equal ($N_\mathrm{p}=N_{\mathrm{p},\,A}=N_{\mathrm{p},\,B}$), the minimum free energy is located at $\mu=\partial f/\partial\phi_A=0$, or
\begin{equation} \label{eq:fh_min}
  \chi N_\mathrm{p}=\frac{\ln{[(1-\phi_A)/\phi_A]}}{1-2\phi_A}.
\end{equation}

In the GCM, the $\chi$ parameter can be modeled by the excess repulsion parameter $\Delta A=A_{AB}-A_{AA}$, with $A_{AA}=A_{BB}$. As a first step to map $\Delta A$ to the $\chi$ parameter, we ran a series of MD simulations of $A$ and $B$ monomers in the canonical ($NVT$) ensemble at temperature $T=300$ $\mathrm{K}$ and varying $\Delta A$ values, with each system having dimensions of $20d\times20d\times50d$ and $N_A=N_B=25,000$ particles of each type. The initial macrophase-separated configurations were generated by randomly placing only $A$ or $B$ particles in the left and right halves of the simulation box, respectively. For each simulation, energy minimization and system equilibration were performed over $5\times10^6$ timesteps of step size $t=0.01\tau$, and data was collected over at least an additional $2\times10^7$ timesteps, which is on the order of $630$ $\mathrm{ns}$.

\begin{figure}[h]
\centering
  \includegraphics[width=3.25in]{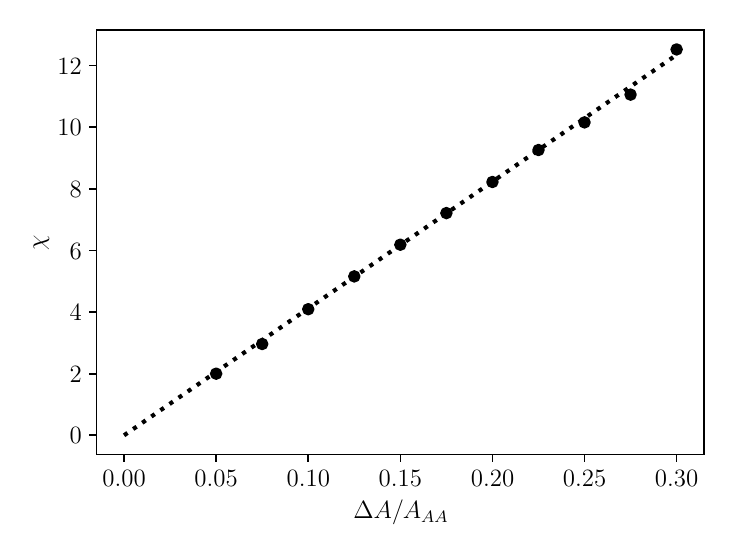}
  \caption{The Flory--Huggins $\chi$ parameter as a function of the excess repulsion parameter $\Delta A$ for systems with a number density of $\rho=2.5d^{-3}$. The dotted line show the best linear fit.}
  \label{fig:water_fh_mapping}
\end{figure}

\begin{figure}[H]
\centering
  \includegraphics[width=3.25in]{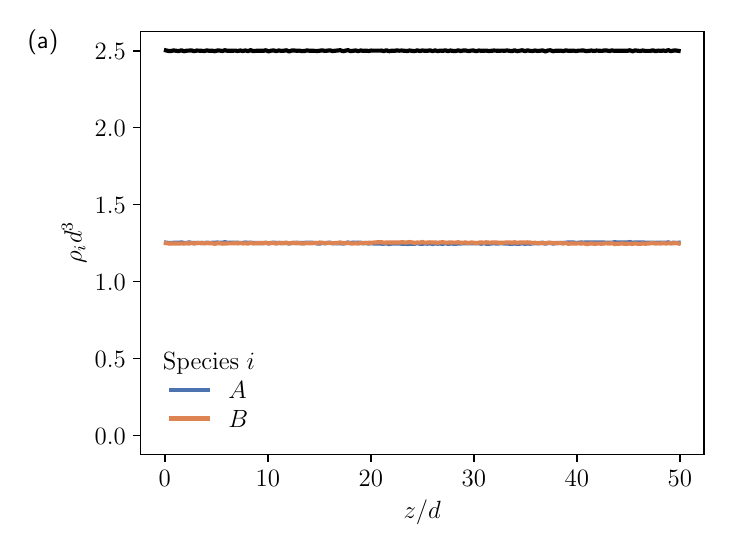}
  \includegraphics[width=3.25in]{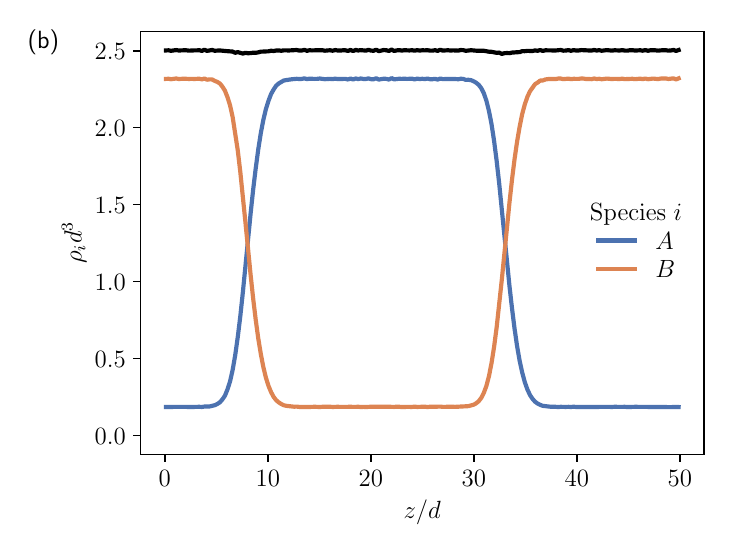}
  \includegraphics[width=3.25in]{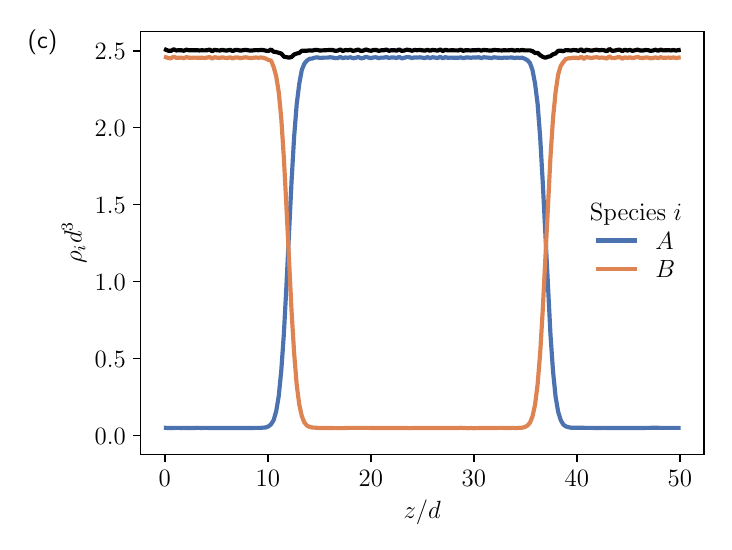}
  \caption{Number density profiles $\rho_i(z)$ of the $A$ and $B$ particles at $\Delta A/A_{AA}$ values of (a) $0.050$, (b) $0.075$, and (c) $0.100$.}
  \label{fig:si_water_fh_number_density}
\end{figure}

To determine the $\chi$--$\Delta A$ relationship, we first computed the number density profiles of $A$ and $B$ particles across the interface by binning the $z$-axis into intervals, counting the number of relevant particles, and dividing the counts by the volume of the bins. Selected number density profiles are shown in Fig. \ref{fig:si_water_fh_number_density}. Then, the volume fraction $\phi_A=\rho_A/(\rho_A+\rho_B)$ for each $\Delta A$ value was evaluated using the average values of $\rho_A$ and $\rho_B$ over the $A$-rich domain where the total number density is homogeneous. Finally, the $\chi$ values were calculated by substituting the $\phi_A$ values into Eq. \ref{eq:fh_min}. Fig. \ref{fig:water_fh_mapping} shows that $\chi$ is linearly proportional to $\Delta A$, and linear regression over the intermediate $\Delta A$ regime yields the explicit scaling
\begin{equation} \label{eq:fh_mapping}
  \chi=41.15\Delta A/A_{AA}=1.63\Delta A/(N_\mathrm{A}d^3k_\mathrm{B}T)
\end{equation}
for systems with a number density of $\rho=2.5d^{-3}$. As such, Eq. \ref{eq:fh_mapping} is an effective mapping of the GCM onto the FH theory.

\section*{Illustrative example}
\subsection*{Electrostatic potential profiles}
The electrostatic potential profile $\Psi(z)$ can determined by numerically evaluating the one-dimensional Poisson's equation
\begin{equation}
  \varepsilon_0\varepsilon_\mathrm{r}\frac{\partial^2\Psi}{\partial z^2}=-\rho_q(z)
\end{equation}
using the charge density profile $\rho_q(z)$ with boundary conditions $\left.\Psi\right|_{z=0}=0$ to use the left electrode as the reference electrode and $\left.\partial\Psi/\partial z\right|_{z=0}=-\sigma_q/(\varepsilon_0\varepsilon_\mathrm{r})$ to enforce zero electric field in the bulk of the polyelectrolyte system.\cite{bagchi_surface_2020,boda_calculating_2021} Without an applied potential difference ($\Delta V=0)$, the surface charge density $\sigma_q$ is zero in systems with nonmetal electrodes and $\sigma_q=-L_z^{-1}\int_0^{L_z}z\rho_q(z)\,dz$ in systems with perfectly conducting electrodes due to the polarization effects from the image charges.\cite{hautman_molecular_1989,qing_surface_2021}

\begin{figure}
\centering
  \begin{subfigure}{3.226in}
  \includegraphics[width=3.226in]{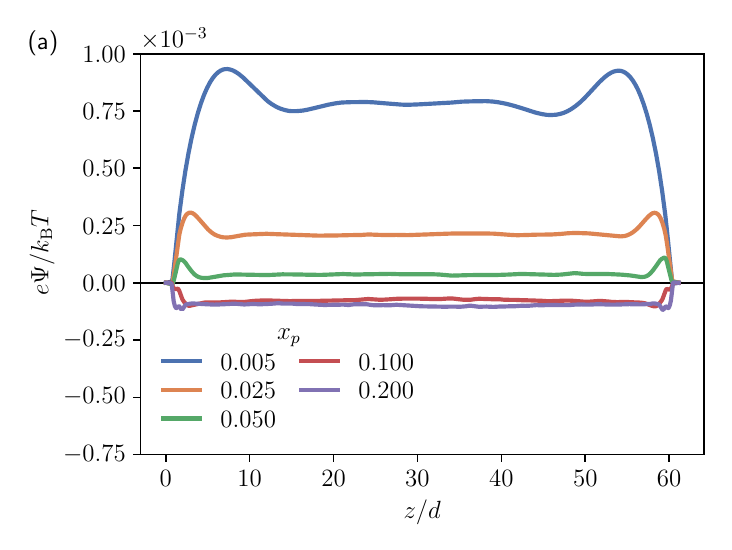}
  \end{subfigure}
  \begin{subfigure}{3.226in}
  \includegraphics[width=3.226in]{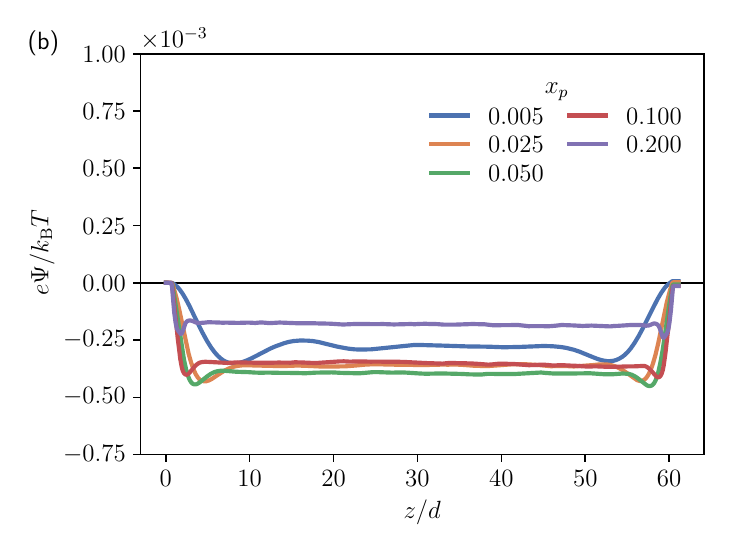}
  \end{subfigure}
  \caption{Electrostatic potential profiles $\Psi(z)$ for systems with (a) perfectly conducting and (b) nonmetal electrodes at different ion number fractions $x_\mathrm{p}$.}
  \label{fig:si_pcs_potential}
\end{figure}

\bibliography{bibliography}